\documentclass[10pt,twocolumn,twoside,journal]{IEEEtran}
\IEEEoverridecommandlockouts
\usepackage{subfigure} 
\usepackage{graphicx}

\usepackage{amsmath,graphicx,amssymb,mathtools,bm}
\usepackage{subfigure}
\usepackage{hyperref}
\usepackage{cite}
\usepackage{amsmath,amssymb,amsfonts}
\usepackage{algorithmic}
\usepackage{graphicx}
\usepackage{textcomp}
\usepackage{xcolor}
\usepackage{verbatim}  
\usepackage{graphicx}  
\usepackage{bm}  
\usepackage{mathrsfs} 
\usepackage{algorithm} 
\usepackage{algorithmic} 
\usepackage{booktabs}
\usepackage{textcomp}  
\usepackage{multirow}  
\usepackage{lettrine}   
\usepackage{color}  
\usepackage{amsmath}
\usepackage{amssymb}
\usepackage{stfloats} 
\usepackage{caption}
\usepackage{subfigure}

\usepackage{color}  
\usepackage[linesnumbered,ruled,vlined,algo2e]{algorithm2e} 
\UseRawInputEncoding

\begin{document}
	\newcommand{\tabincell}[2]{\begin{tabular}{@{}#1@{}}#2\end{tabular}}


\title{Extremely Large-Scale Dynamic Metasurface Antennas (XL-DMAs): Near-Field Modeling and Channel Estimation
}

\author{Songjie Yang, Wanting Lyu, Boyu Ning, \IEEEmembership{Member,~IEEE}, Yue Xiu, Youzhi Xiong, \IEEEmembership{Member,~IEEE},\\ Hua Chen, \IEEEmembership{Senior Member,~IEEE}, Chadi Assi, \IEEEmembership{Fellow,~IEEE}, and Chau Yuen, \IEEEmembership{Fellow,~IEEE}
	
	\thanks{Songjie Yang, Wanting Lyu, and Boyu Ning are with the National Key Laboratory of Wireless Communications, University of Electronic Science and Technology of China, Chengdu 611731, China. 
		(e-mail:
		yangsongjie@std.uestc.edu.cn;
		lyuwanting@yeah.net; boydning@outlook.com).
		
		Yue Xiu is with College of Air Traffic Management, Civil Aviation Flight University of China, Sichuan, China, 618311. (e-mail: xiuyue12345678@163.com).
		
		Youzhi Xiong is with College of Physics and Electronic Engineering,
		Sichuan Normal University, Chengdu 610101, China. (e-mail: yzxiong@sicnu.edu.cn).
		
		Hua Chen is with the Faculty of Electrical Engineering
		and Computer Science, Ningbo University, Ningbo 315211, China. (e-mail:
		dkchenhua0714@hotmail.com).
		Hua Chen is also with the Zhejiang Key Laboratory of Mobile Network
		Application Technology, Ningbo 315211, P. R. China.
		
		Chadi Assi is with Concordia University, Montreal, Quebec, H3G 1M8, Canada (email:assi@ciise.concordia.ca).
		
		Chau Yuen is with the School of Electrical and Electronics Engineering, Nanyang Technological University (e-mail: chau.yuen@ntu.edu.sg).}}
\maketitle

\begin{abstract}
	Dynamic metasurface antennas (DMAs) represent a novel transceiver array architecture for extremely large-scale (XL) communications, offering the advantages of reduced power consumption and lower hardware costs compared to conventional arrays.
	 This paper focuses on near-field channel estimation for XL-DMAs. We begin by analyzing the near-field characteristics of uniform planar arrays (UPAs) and introducing the Oblong Approx. model. This model decouples elevation-azimuth (EL-AZ) parameters for XL-DMAs, providing an effective means to characterize the near-field effect. It offers simpler mathematical expressions than the second-order Taylor expansion model, all while maintaining negligible model errors for oblong-shaped arrays.
	Building on the Oblong Approx. model, we propose an EL-AZ-decoupled estimation framework that involves near- and far-field parameter estimation for AZ/EL and EL/AZ directions, respectively. The former is formulated as a distributed compressive sensing problem, addressed using the proposed off-grid distributed orthogonal least squares algorithm, while the latter involves a straightforward parallelizable search. Crucially, we illustrate the viability of decoupled EL-AZ estimation for near-field UPAs, exhibiting commendable performance and linear complexity correlated with the number of metasurface elements.
	Moreover, we design an measurement matrix optimization method with the Lorentzian constraint on DMAs and highlight the estimation performance degradation resulting from this constraint.
\end{abstract}
\section{Introduction}  
Researchers from both industry and academia are actively engaged in defining the features and technologies for the forthcoming 6G networks. Among the anticipated attributes is its capacity to accommodate a wide spectrum of applications, with a notable focus on facilitating ultra-high-speed communication and ultra-high-resolution sensing capabilities. Meeting these requirements necessitates the use of high-frequency technologies like millimeter-wave and terahertz, in conjunction with a larger number of antennas to yield substantial beamforming gains and achieve remarkable spatial resolution.

During the transition from multi-input multi-output (MIMO) to massive MIMO, significant structural enhancements have been implemented to overcome the cost and energy consumption challenges arising from the expanded number of antennas. For example, the originally deployed fully-digital structure, in which each antenna is connected to one radio frequency (RF) chain,
 proved inadequate for massive MIMO scenarios. Consequently, two distinct approaches have emerged to address this issue: the utilization of cost-effective hardware solutions, including one-bit analog-to-digital converters, and the development of hybrid analog-digital beamforming architectures, both of which effectively contribute to reducing overall system costs.
 Furthermore, these approaches have brought about significant shifts in the landscape of array signal processing for wireless communications, ushering in a paradigm shift. Recent research endeavors, centered around the utilization of hybrid beamforming architectures and harnessing the sparsity inherent in high-frequency channels, have led to notable advancements in various fields, encompassing compressive/sparse channel estimation \cite{CE1,CE2,CE3,CE4}, beam training \cite{BT1,BT2,BT3}, and hybrid beamforming \cite{HBF1,HBF2}.
 
 The transition within MIMO technology continues to evolve, with recent attention focusing on extremely large-scale (XL)-MIMO systems for high-frequency communications, offering enhancements in both sensing and communication capabilities. This technique introduces the near-field effect, prioritizing spherical-wave propagation over planar-wave propagation in the near-field region. This region is determined by the Fraunhofer distance, also known as the Rayleigh distance, which represents the minimum distance required to maintain a phase difference of received signals across the array elements at a maximum of $\pi/8$.
 The increased Fraunhofer distance, a result of the large aperture of XL-arrays, should be carefully considered within the typical communication coverage range due to its undeniable impact. Consequently, it challenges the conventional far-field assumption employing planar wavefronts used in traditional communications. In response, researchers have explored near-field signal processing, including channel estimation \cite{NF-CE1,NF-CE3,NF-CE2,NF-CE4,NF-CE5,NF-CE6,NF-CE7}, sensing \cite{LOC3,LOC1,LOC2}, and hybrid beamforming \cite{NFHBF}.
 Those challenges are accompanied by significant opportunities. Leveraging the near-field effect, XL-arrays can unlock additional benefits that extend beyond their substantial beamforming gains, with a particular focus on line-of-sight multiplexing and capacity enhancement \cite{LOS-MIMO2}.

In line with the evolution of XL-MIMO, there arises a crucial demand for novel array structural technologies adept at accommodating the increased number of antennas inherent in XL-MIMO systems. The advent of metamaterials, which are engineered composite materials capable of exhibiting unique electromagnetic properties, opens up new avenues for array antennas. In recent years, there has been a notable upswing in the exploration of metasurfaces as reflective elements in wireless communication systems. In such scenarios, reflecting surfaces are adjustable to improve the wireless channel and provide more degrees-of-freedom \cite{RIS1,RIS2,RIS3,RIS4,RIS5,RIS6}.  
 Additionally, there is a growing trend in wireless communications to leverage metasurfaces for their ability to control radiation and reception patterns. This transformation turns metasurfaces into active transceiving devices, departing from their traditional role as passive reflectors. Dynamic metasurface antennas (DMAs), a form of reconfigurable antenna, exemplify this trend. They operate by adjusting the resonant frequency of individual elements to achieve the desired radiated amplitude and phase for beamforming. This array antenna typically exhibits considerably lower power consumption and cost compared to standard arrays \cite{DMA0,DMA00}. 
 Therefore, the use of DMA-based transceivers shows great promise for XL-MIMO, as they require significantly less power and cost compared to conventional phased-array counterparts.

 The research on DMAs for wireless communications is in its infancy. One key challenge that sets DMAs apart from phased arrays is the Lorentzian constraint (LC) imposed on element weights \cite{META1,META2}. This constraint enforces that all phases in the range of $[0, \pi]$ have distinct, yet non-adjustable, amplitudes. 
  In \cite{DMA1}, a mathematical model was proposed for massive MIMO systems with DMAs and discuss their constraints compared to standard antenna arrays, which  revealed
 the potential advantages of using DMAs over standard arrays for realizing low-power massive MIMO. Considering the practical hardware constraint, the authors of \cite{DMA2} jointly optimized the DMA weights along with the dynamic range of the ADCs
 and the digital processing, under a given bit constraint.
Most of the
existing works focused on DMA-based spectral efficiency optimization, while
DMA-based energy efficiency optimization has rarely
been explored.  \cite{DMA3} filled this gap.
Unlike the previous studies which relaxed the LC on the DMA elements, the authors of \cite{DMA4} studied downlink beamforming for DMAs without employing
performance-degrading relaxations. All of the aforementioned research has primarily concentrated on the far field, while the exploration of near-field DMAs has gained attention in works such as \cite{NFDMA1, NFDMA2}, particularly emphasizing downlink beam focusing and localization.

So far, there has been a noticeable absence of near-field channel estimation research for DMAs. Considering the crucial role that accurate channel state information plays in supporting the aforementioned studies, further exploration in this area is imperative. It is noteworthy that research on channel estimation errors stemming from the LC in comparison to ideal array configurations is also lacking. Moreover, it is equally crucial to acknowledge that DMAs serve as a pivotal array configuration enabling XL-MIMO. 
 These considerations motivate us to study channel estimation for XL-DMAs in this paper. 
  While our primary focus is on near-field XL-DMAs, the proposed estimation method can also be applied to conventional far-field DMAs. We summarize our primary contributions and innovations as follows:

\begin{itemize}
	\item \textbf{Near-Field XL-DMA Modeling:} First, we embark on an analysis of the near-field XL-DMA model, delving into the intricacies it presents when processing near-field uniform planar arrays (UPAs) in contrast to their far-field counterparts and near-field uniform linear arrays (ULAs). To enhance the practicality of signal processing, we introduce a model approximation referred to as Oblong Approx. which effectively decouples elevation-azimuth (EL-AZ) parameters via a Kronecker product.
	The essence of Oblong Approx. lies in the recognition that employing oblong-shaped arrays, where one side's length significantly exceeds the other, induces a near-field effect that can be safely neglected on the shorter side. This concept seamlessly integrates with the XL-DMA structure, characterized by the incorporation of several RF chains (assumed in EL), each connected to an extensive microstrip (assumed in AZ).
	Following this, we employ diverse metrics to assess the model error of Oblong Approx. as we increase the number of RF chains.
	\item \textbf{EL-AZ-DE Framework:}  
	 Notably, the Oblong Approx. model highlights that the AZ/EL direction primarily emphasizes the far-field parameter, in contrast to the near-field EL/AZ parameter. In this context, we develop an EL-AZ-decoupled estimation (short for EL-AZ-DE) framework that initially integrates all RF chains to jointly estimate AZ distance-angle parameters for each microstrip. Subsequently, it employs the Kronecker product structure to perform EL angle estimation.
	\item \textbf{Distributed Super-Resolution  Recovery:} The joint estimation for the AZ direction falls within the realm of a distributed compressive sensing problem. To address this and account for dictionary mismatch resulting from limitations in recovering true parameters due to finite dictionary resolution, we devise an off-grid distributed orthogonal least squares (OG-DOLS) algorithm. This algorithm adeptly integrates on-grid distance-angle parameter selection with off-grid distance-angle parameter refinement across multiple sensing matrices. Moreover, we propose measurement matrix optimization (MMO), also known as beam pilot optimization, within the LC for improved estimation.
	  
	\item \textbf{DMAs vs. Phased-Arrays:} As this paper marks the inaugural study on channel estimation for DMAs, we assess the impact of the LC on channel estimation by contrasting it with the ideal phased-array counterpart.
\end{itemize}
 
The rest of this paper is organized as follows: Section \ref{Sys} describes the DMA model, uplink training model, and the channel model.
Section \ref{OBA} establishes the Oblong Approx. model for
XL-DMAs, along with a model error assessment. Section \ref{NFCE} explores efficient channel estimation frameworks for XL-DMAs. Section \ref{MMO} proposes the MMO method under the LC. 
 In Section \ref{simu}, we perform various numerical simulations to evaluate the effectiveness of the proposed methods. 
Finally, Section \ref{Con} concludes this paper.

{\emph {Notations}}:
${\left(  \cdot  \right)}^{ *}$, ${\left(  \cdot  \right)}^{ T}$, ${\left(  \cdot  \right)}^{ H}$, and $\left(\cdot\right)^{-1}$ denote conjugate, transpose, conjugate transpose, and inverse, respectively.  $\Vert\cdot\Vert_0$ and $\Vert\cdot\Vert_2$ represent $\ell_0$ norm and $\ell_2$ norm, respectively. 
$\Vert\mathbf{A}\Vert_F$ denotes the Frobenius norm of matrix $\mathbf{A}$. ${\rm Tr}\{\mathbf{A}\}$ denotes the trace of matrix $\mathbf{A}$. $\vert\cdot\vert$ denotes the modulus. Furthermore, $\otimes$ is the Kronecker product. $[\mathbf{a}]_{i}$ and $[\mathbf{A}]_{i,j}$ denote the $i$-th element of vector $\mathbf{a}$, the $(i,j)$-th element of matrix $\mathbf{A}$, respectively. $[\mathbf{A}]_{i,:}$ and $[\mathbf{A}]_{:,j}$ denote the $i$-the row and the $j$-the column of matrix $\mathbf{A}$, respectively. $\rm{vec}(\cdot)$ represents the vectorization operation. $\mathbb{E}\{\cdot
\}$ denotes the expectation operations.  $\mathbf{I}_M$ denotes the $M$-by-$M$ identity matrix. Moreover, $\bullet$ is the dot product. Finally, $\mathcal{CN}(\mathbf{a},\mathbf{A})$ is the complex Gaussian distribution with mean $\mathbf{a}$ and covariance matrix $\mathbf{A}$.
 
\begin{figure*}
	\centering
	\includegraphics[width = 0.87\textwidth]{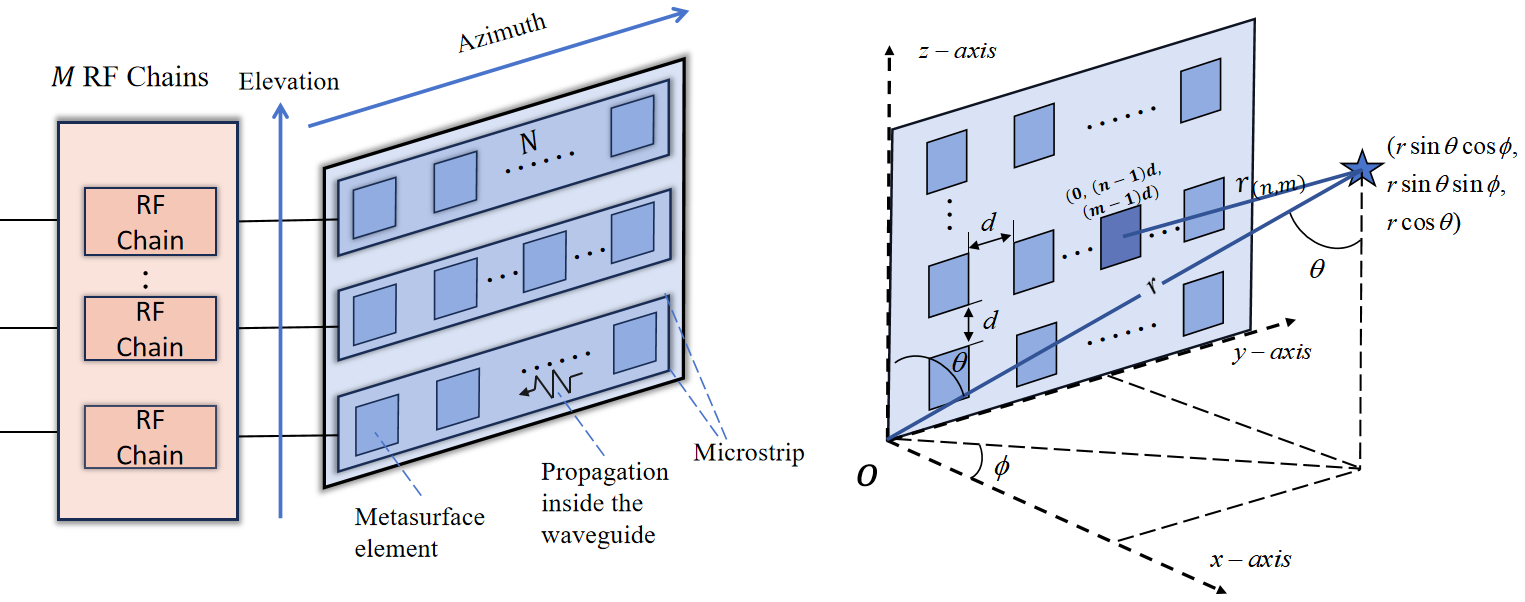}
	\caption{The DMA structure and its three-dimensional coordinate system model.}
	\label{sys_fig}
\end{figure*} 
\section{System Model}\label{Sys}
We consider a near-field uplink narrowband training scenario in which the base station utilizes an XL-DMA with $M$ RF chains, each connected to a microstrip containing $N$ metasurface elements. The users in this scenario are equipped with a single antenna. 

\subsection{DMA Model}
DMAs consist of a multitude of reconfigurable metamaterial
radiating elements that can be used both as transmit and
receive antennas. Those elements are placed on a waveguide
through which the signals to be transmitted, and the received
waveforms intended for information decoding, are transferred.
The transceiver digital processor, which generates the transmit
signals and processes the received signals, is connected to the
waveguide through dedicated input and output ports, respectively.
Due to the implementation difficulty of the two-dimensional waveguide, multiple one-dimensional waveguides for DMAs are commonly used based on microstrips \cite{DMA0}, as shown in Fig. \ref{sys_fig}. 

\begin{figure}
	\centering
	\includegraphics[width = 0.474\textwidth]{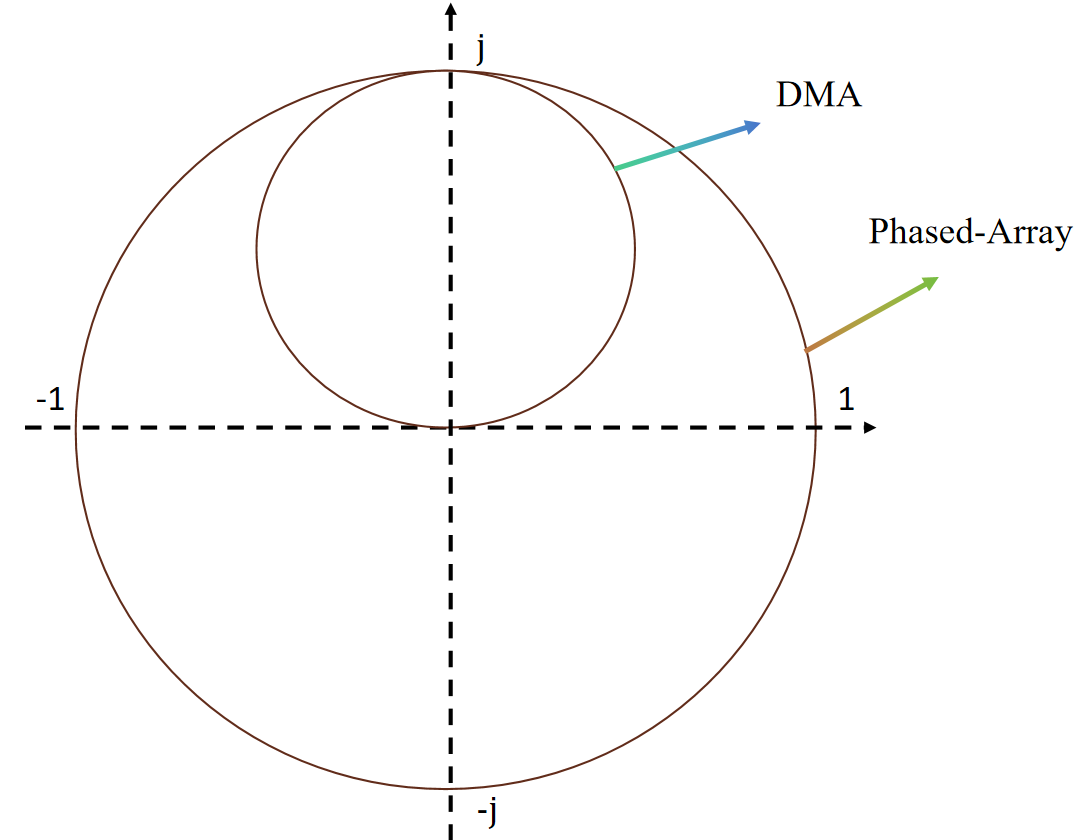}
	\caption{ The feasible range of weights for DMAs and phased-arrays.}
	\label{DMAPA}
\end{figure}

Generally, DMAs exhibit two important physical characteristics: 1) frequency response of the metamaterial element, and 2) propagation inside the waveguide. Since each DMA radiating element acts as a resonant electrical circuit, it results in frequency-selective responses. For simplicity, this paper considers flat frequency responses with the narrowband system assumed. In this case, the state configuration of each DMA element with the LC takes the following form:
\begin{equation}
	q\in\mathcal{Q}\triangleq\left\{\frac{j+e^{jx}}{2}| x\in[0,2\pi]\right\},
\end{equation}
where $x$ represents the configurable phase shift. The feasible range of weights for both DMAs and phased-arrays is illustrated in Fig. \ref{DMAPA}.

Moreover, the propagation inside the waveguide undergoes  attenuation and
phase shift, depending on the location of the elements
along the waveguide. This effect on the $n$-th element of microstrip $m$ can be expressed by
\begin{equation}
	v_{m,n}=e^{-\rho_{m,n}(\alpha_m+j\beta_m)}, \forall m,n,
\end{equation}
where $\rho_{m,n}$, $\alpha_m$, and $\beta_m$ denote the location of the $n$-th element on microstrip $m$, the attenuation coefficient, and the
wavenumber of microstrip $m$, respectively.

\subsection{Uplink Training}
 Utilizing an orthogonal pilot sequence, the process of multi-user channel estimation can be viewed as a collection of parallelizable tasks. For the sake of clarity and without loss of generality, we will discuss the pilot signal model of an arbitrary user.
The received signal at the $m$-the RF chain for the $p$-th pilot symbol ($m\in\{1,\cdots,M\}$, $p\in\{1,\cdots,P\}$) is given by
\begin{equation}
	{y}_{m,p}=\mathbf{q}_{m,p}^H\mathbf{V}_m\mathbf{h}_m{s}_p+\mathbf{q}_{m,p}^H\mathbf{V}_m\mathbf{n}_{m,p},
\end{equation}
where $\mathbf{q}_{m,p}\in\mathbb{C}^{N\times 1}$ containing $N$ DMA phase coefficients denotes the $m$-th RF chain's beam for the $p$-the pilot, and $\mathbf{V}_m\triangleq {\rm diag}(v_{m,1},\cdots,v_{m,N})$.
\begin{equation}
	{y}_{m,p}=\mathbf{w}_{m,p}^H\mathbf{h}_m{s}_p+\mathbf{w}_{m,p}^H\mathbf{n}_{m,p},
\end{equation}
where ${s}_p$, set to $1$, is the $p$-th pilot signal, $\mathbf{h}_m\in\mathbb{C}^{N\times 1}$ denotes the wireless channel between the $m$-th RF chain and the user, $\mathbf{w}_{m,p}\in\mathbb{C}^{N\times 1}$ is the combining/measurement vector for channel sampling,
and $\mathbf{n}_{m,p}\in\mathbb{C}^{N\times 1}$ is the independent and identically distributed additive white Gaussian noise following the distribution $\mathcal{C} \mathcal{N}\left(0, \sigma_n^{2}\right)$.

By collecting the $P$ pilots,   the $m$-th RF chain's received signal $\mathbf{y}_m=[y_{m,1},\cdots,y_{m,P}]^T\in\mathbb{C}^{P\times 1}$ is expressed as
\begin{equation}\label{y}
\mathbf{y}_m=\mathbf{W}_m^H\mathbf{h}_m+\widetilde{\mathbf{n}}_m
\end{equation}
where $\mathbf{W}_m\triangleq[\mathbf{w}_{m,1},\cdots,\mathbf{w}_{m,P}]=\mathbf{V}^H_m\mathbf{Q}_m\in\mathbb{C}^{N\times P}$, $\mathbf{Q}_m\triangleq[\mathbf{q}_{m,1},\cdots,\mathbf{q}_{m,P}]$ and $\widetilde{\mathbf{n}}_m\triangleq[\mathbf{n}_{m,1}^H\mathbf{w}_{m,1},\cdots,\mathbf{n}_{m,P}^H\mathbf{w}_{m,P}]^H\in\mathbb{C}^{P\times 1}$.

\subsection{Channel Model}
Next, we specify the physical channel which can characterize
the near-field geometrical structure and limited scattering nature. The channel $\overline{\mathbf{h}}=[\mathbf{h}_1^T,\cdots,\mathbf{h}_M^T]^T\in\mathbb{C}^{MN\times1}$ is written as
\begin{equation}\label{H}
	\begin{aligned}
		\overline{\mathbf{h}}=&\sqrt{\frac{MN}{L}}\sum_{l=1}^{L}z_l\mathbf{g}(\theta_l,\phi_l,r_l)\\
		=& \mathbf{G}\mathbf{z},
	\end{aligned}
\end{equation}
where $L$ is the number of channel paths, $z_l$ denotes the complex gain of the $l$-th path, $\mathbf{G}\triangleq[\mathbf{g}(\theta_1,\phi_1,r_1),\cdots,\mathbf{g}(\theta_L,\phi_L,r_L)]\in\mathbb{C}^{MN\times L}$, $\mathbf{z}\triangleq\sqrt{\frac{MN}{L}}[z_1,\cdots,z_L]^T$, and $\mathbf{g}(\theta_{l},\phi_l,r_l)\in\mathbb{C}^{MN\times 1}$ represents the array manifold following 
\begin{equation}\label{g}
	\begin{aligned}
	&	\mathbf{g}(\theta,\phi,r)\triangleq	  \\ & \frac{1}{\sqrt{MN}}
		\left[ e^{-j\frac{2\pi}{\lambda} (r_{(1,1)}-r)},\cdots,e^{-j\frac{2\pi}{\lambda} (r_{(M,N)}-r)} \right]^T,
	\end{aligned}
\end{equation}
where $r_{(n,m)}$ is shown in Eqn. (\ref{rd}), derived by calculating the distance between the $(m,n)$-th element with vector coordinate $(0,(n-1)d,(m-1)d)$ and the object with vector coordinate $(r\sin(\theta)\cos(\phi),r\sin(\theta)\sin(\phi),r\cos(\theta))$, as illustrated in Fig. \ref{sys_fig}. 
\begin{figure*} 
\begin{equation}\label{rd}
	\begin{aligned}
		r_{(n,m)}  &\triangleq\left((r\sin(\theta)\cos(\phi))^2+(r\sin(\theta)\sin(\phi)-(n-1)d)^2 
		+(r\cos(\theta)-(m-1)d)^2 \right)^{\frac{1}{2}} \\
		&= \sqrt{r^2-2r(n-1)d\sin(\theta)\sin(\phi)+((n-1)d)^2-2r(m-1)d\cos(\theta)+((m-1)d)^2} \\
		& \overset{(a)}{\approx} r-(n-1)d \sin(\theta)\sin(\phi) - (m-1)d\cos(\theta) +\frac{((m-1)d)^2}{2r}(1-\cos^2(\theta))+\frac{((n-1)d)^2}{2r}(1-\sin^2(\theta)\sin^2(\phi)) \\
		& \ \ \ \ 	 -\frac{(n-1)(m-1)d^2\sin(\theta)\sin(\phi)\cos(\theta)}{r} + ...,
	\end{aligned}
\end{equation}
\end{figure*}
where $(a)$ holds due to the second-order Taylor expansion $\sqrt{1+x+y}\approx1+\frac{x+y}{2}-\frac{(x+y)^2}{8}$.

\begin{figure*}
	\centering
	\includegraphics[width = 0.87\textwidth]{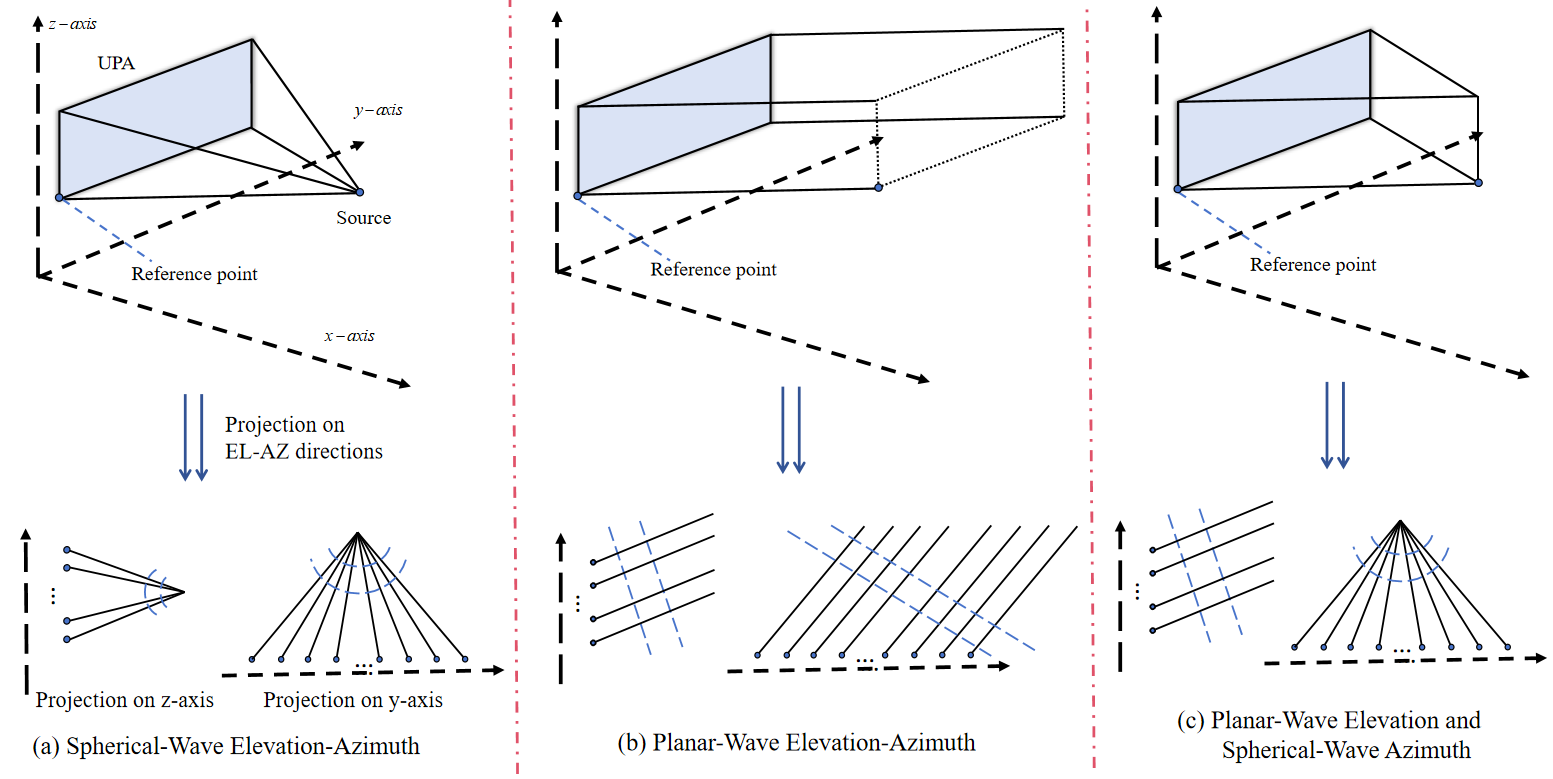}
	\caption{Three kinds of propagation models in EL-AZ directions. }
	\label{EMP}
\end{figure*} 

\section{Proposed Oblong Approx. Model for XL-DMAs}\label{OBA}
Processing the spherical-wave array manifold directly poses challenges, leading to a commonly used approximation method involving the second Taylor expansion to expand $r_{(m,n)}$ as detailed in Eqn. (\ref{rd}). This expansion simplifies $r_{(m,n)}$ into a sum of terms, none of which exceed quadratic. However, this demonstrates that the near-field array manifold relies not only on coupled angle-distance terms but also on coupled EL-AZ terms.
Alternatively, the second-order Taylor approximation for UPAs remains intricate due to the cross term $\frac{nmd^2\sin(\theta)\sin(\phi)\cos(\theta)}{r}$. Recognizing the physical structure of XL-DMAs, which deploy a significantly larger number of elements along the $y$-axis compared to the $z$-axis, we propose a novel approximation tailored for XL-DMAs. This approach effectively decouples the AZ and EL directions and assumes planar-wave and spherical-wave models for each direction, respectively.
 This suggests
\begin{equation}\label{gD}
	\mathbf{g}(\theta,\phi,r)\approx\mathbf{g}_D(\theta,\phi,r)\triangleq \mathbf{a}(\vartheta)\otimes \mathbf{b}(\varphi,r),
\end{equation} 
where $\vartheta\triangleq \cos(\theta)$ and $\varphi\triangleq \sin(\theta)\cos(\phi)$.  $\mathbf{a}(\vartheta)\in\mathbb{C}^{M\times 1}$ and $\mathbf{b}(\varphi,r)\in\mathbb{C}^{N\times 1}$ are given by
\begin{equation}
\mathbf{a}(\vartheta)\triangleq\sqrt{\frac{1}{M}}\left[1, e^{j\frac{2\pi}{\lambda}d\vartheta},\cdots, e^{j\frac{2\pi}{\lambda}(M-1)d\vartheta}\right]^T,
\end{equation}
and
\begin{equation}
	\begin{aligned}
	&\mathbf{b}(\varphi,r)\triangleq\sqrt{\frac{1}{N}}\left[1, e^{j\frac{2\pi}{\lambda}\left(d\varphi-\frac{d^2}{2r}(1-\varphi^2)\right)},\cdots, \right. \\ &\ \ \ \ \ \ \ \ \ \ \ \ \ \ \ \ \ \ \left. e^{j\frac{2\pi}{\lambda}\left((N-1)d\varphi-\frac{((N-1)d)^2}{2r}(1-\varphi^2)\right)}\right]^T.
	\end{aligned}
\end{equation}

We show in Table \ref{TAB} the element distance of different array manifold model. First, a significant phase transition is evident when moving from ULA to UPA using the second-order Taylor approximation. In contrast, under the far-field assumption, this transition is more pronounced. Additionally, it is notable that while the spherical wavefront model can be applied to both near- and far-field regions, its mathematical expression makes it challenging to extract spatial information effectively. On the other hand, the second-order Taylor approximation provides a more transparent insight into how angles and distances govern the array manifold. Nevertheless, it introduces complexity owing to the coupled $\varphi$-$\vartheta$ term. In contrast, the proposed Oblong Approx. is notably more straightforward in the UPA scenario due to its further simplification.
Furthermore, the planar wavefront model boasts the simplest expression but suffers from significant phase loss in the near-field region.

\subsection{Intuitive Insights}
Here, we begin by offering some intuitive insights into the rationale behind Eqn. (\ref{gD}). Since each RF chain is connected to an analog ULA microstrip, it is inherently limited to sensing a single spatial direction, which, in our scenario, corresponds exclusively to the AZ direction. Consequently, for DMAs, capturing information about the EL direction necessitates the involvement of two or more RF chains. In essence, the analog end of the DMA system senses the AZ direction, while the baseband end is responsible for capturing signals from the EL direction. In light of this hardware configuration, a heuristic approach involves aligning the analog and baseband ends with the AZ and EL directions of the channel, respectively, as illustrated in the left hand of Fig. \ref{sys_fig}.

On the other hand, the quantity of RF chains is typically less than the number of antenna elements in XL-DMAs, signifying that there are fewer elements along the EL compared to the AZ direction. In this sense, a question naturally arises---\emph{Is it permissible to treat the wavefronts of electromagnetic waves in the EL and AZ directions with bias or unfairness?} To address this query, we delve deeper into the second-order Taylor expansion and notice that the distance parameter is associated with quadratic terms concerning $m$ or $n$, indicating that the EL direction has a comparatively smaller impact on the distance component, given that $m^2$ and $mn$ are significantly smaller than $n^2$. Therefore, by disregarding the terms concerning  $m^2$ and $mn$, Eqn. (\ref{gD}) is obtained.
Finally, we establish a cross-field model that employs the spherical- and planar-wave propagation for the AZ and EL directions, respectively. This is visually shown in Fig. \ref{EMP}(c). Compared with Figs. \ref{EMP}(a) and (b) that show the spherical- and planar-wave propagation, respectively, Fig. \ref{EMP}(c) makes a trade-off to simplify the near-field UPA model.

\subsection{Model Evaluation}

The above presentation offers an intuitive depiction of our proposed model approximation. In this context, we assess the model's accuracy from two metrics: element distance and beamforming.

\begin{table*}
	\caption{Element Distance Parameter Comparison using ULAs and UPAs in Different Propagation Models}
	\label{TAB}
	\centering
	\begin{tabular}{ |c|c|| c | c |  }
		\hline
		\tabincell{c}{Array Layout}&Propagation Model &  Element Distance $r_{(m,n)}$ & \tabincell{c}{Applicable Region}  \\ \hline\hline
		\multirow{4}{*}{\tabincell{c}{\tabincell{c}{
					ULA\\ (along $y$-axis)} }} & Spherical Wavefront & $\sqrt{r^2-2r(n-1)d\varphi+((n-1)d)^2}$ &Near to far field \\ \cline{2-4}
		& Second-order Taylor Approx.& $r-(n-1)d \varphi  +\frac{((n-1)d)^2}{2r}(1-\varphi^2)  $&Near field\\
		\cline{2-4}
		&Proposed Oblong Approx. & $r-(n-1)d \varphi  +\frac{((n-1)d)^2}{2r}(1-\varphi^2)$&Near field\\
		\cline{2-4}
		&\tabincell{c}{Planar Wavefront\\ (Far-Field Assumption)}   & $r-(n-1)d \varphi   $&Far field\\
		\cline{2-4}

		\hline
		\multirow{4}{*}{\tabincell{c}{\tabincell{c}{UPA\\ (along $y$-$z$-plane)}}} & Spherical Wavefront & $\sqrt{r^2-2r(n-1)d\varphi+((n-1)d)^2-2r(m-1)d\vartheta+((m-1)d)^2}$ &Near to far field \\ \cline{2-4}
		& Second-order Taylor Approx.& \tabincell{c}{$r-(n-1)d \varphi - (m-1)d\vartheta +\frac{((m-1)d)^2}{2r}(1-\vartheta^2)$\\$+\frac{((n-1)d)^2}{2r}(1-\varphi^2) 	 -\frac{(n-1)(m-1)d^2\varphi\vartheta}{r}$}&Near field\\
		\cline{2-4}
		&Proposed Oblong Approx. & $r-(n-1)d \varphi - (m-1)d\vartheta  +\frac{((n-1)d)^2}{2r}(1-\varphi^2) 	 $&Near field\\
		\cline{2-4}
		&\tabincell{c}{Planar Wavefront\\ (Far-Field Assumption)}   & $r-(n-1)d \varphi - (m-1)d\vartheta   	 $&Far field\\
		\cline{2-4}
		
		\hline

	\end{tabular}
\end{table*}
\subsubsection{Element Distance}
 
From Table \ref{TAB}, we can observe the distances of each antenna element with various models, which we previously subjected to a qualitative analysis. Now, we will conduct a quantitative analysis of the UPA part, using the element distance error with respect to other models and the ideal model (Spherical Wavefront) as the metric. The numerical result can be found in Fig. \ref{ED}.

\subsubsection{Beamforming}
For single-path beamforming, the objective is to align with the channel's array manifold, i.e., $\mathbf{g}_D^H\mathbf{g}$. To evaluate the accuracy of the model approximation, different array manifolds are applied to single-path beamforming, and the beamforming gain error is a key indicator. The numerical result can be found in Fig. \ref{BF}.

  \section{Near-Field Channel Estimation for XL-DMAs}\label{NFCE}
  
  This section explores efficient channel estimation frameworks for XL-DMAs, including elevation-azimuth-joint estimation (EL-AZ-JE), azimuth-individual estimation (AZ-IE), and EL-AZ-JE. These frameworks demonstrate how to recover the near-field spatial parameters ${\vartheta,\varphi,r}$ from measured signals and reconstruct the channel.  
  \subsection{EL-AZ-JE}\label{ELAZJE}
 From the perspective of sparse signal processing, the measured signals with the sensing matrix can formulate a direct sparse recovery problem with respect to $\{\vartheta,\varphi,r\}$:
  \begin{equation}\label{DE}
  \begin{aligned}
  & \underset{\vartheta,\varphi,r}{\rm arg \ min} \left\Vert \mathbf{z} \right\Vert_0 \\
  & {\rm s.t.} \ \left\Vert\widetilde{\mathbf{y}}-\widetilde{\mathbf{W}}^H\overline{\mathbf{G}}\mathbf{z}\right\Vert_2^2\leq \epsilon, 
  \end{aligned}
 \tag{$\mathcal{P}_1$}
  \end{equation}
where $\epsilon$ represents the precision factor, $\widetilde{\mathbf{W}}\triangleq{\rm blkdiag}\{\mathbf{W}_1,\cdots,\mathbf{W}_M\}$, $\widetilde{\mathbf{y}}=[\mathbf{y}_1^T,\cdots,\mathbf{y}_M^T]^T$,
 $\overline{\mathbf{G}}\in\mathbb{C}^{MN\times \overline{G}}$ is the angle-distance dictionary constructed from $\mathbf{g}(\theta,\phi,r)$ with different $\{\theta,\phi,r\}$ samples, and $\mathbf{z}$ is a sparse vector where each non-zero element index indicates one $\{\vartheta,\varphi,r\}$ combination.

Problem (\ref{DE}) is a standard compressive sensing problem, which
 can be addressed using various recovery algorithms.
 However, the use of a large-scale dictionary $\widetilde{\mathbf{W}}$ can lead to unacceptable time complexity and adversely affect estimation performance due to dictionary redundancy. As a result, problem (\ref{DE}) is not an ideal choice for near-field estimation.
\subsection{AZ-IE}\label{AZIE}
To avoid the use of a large-scale dictionary, one effective approach is to employ a small-scale angular-domain dictionary for AZ angle estimation for each microstrip. For the $m$-th microstrip, the AZ angle estimation problem can be formulated as follows:
  \begin{equation}\label{IE}
	\begin{aligned}
		& \underset{\varphi,r}{\rm arg \ min}  \left\Vert \bm{\xi}_m\right\Vert_0 \\
		& {\rm s.t.} \   \left\Vert{\mathbf{y}}_m-\mathbf{W}^H_m\mathbf{B}\bm{\xi}_m\right\Vert_2^2\leq \epsilon, 
	\end{aligned}
	\tag{$\mathcal{P}_2$}
\end{equation}
 where $\mathbf{B}\in\mathbb{C}^{N\times \overline{g}}$ is the $\varphi$-$r$ dictionary that represents $\mathbf{h}_m$ in a sparse vector $\bm{\xi}_m$, with each non-zero element indicating one $\{\varphi,r\}$ combination, and $\overline{g}$ is the number of AZ angle-distance samples.
 
This sparse recovery framework provides a low-complexity solution, but it overlooks the direct estimation of the EL parameter, resulting in a degradation of estimation performance.
\subsection{EL-AZ-DE}\label{ELAZDE}
To overcome the limitations of the aforementioned methods, we leverage our proposed Oblong Approx. model to streamline the parameter estimation process. It estimates $\{\varphi, r\}$ first and then proceeds with $\vartheta$. As this estimation process spans from near- to far-field parameters, we refer to it as CFE.
 Referring to Eqs. (\ref{H}) and (\ref{gD}), it is evident that the AZ direction of the channel is consistent across all RF chains. That is, $\{\mathbf{h}_m | m=1, \cdots, M\}$ share the same spatial parameters. Additionally, the Kronecker product in Eqn. (\ref{gD}) indicates that the channel is formed by weighting the EL array manifold with the AZ array manifold. Therefore, once the AZ parameter is known, obtaining the EL parameter becomes a straightforward task. 
  
  \subsubsection{Estimation of $\{\varphi_l,r_l\}_{l=1}^L$}
  As stated before, the signals $\{\mathbf{y}_m\}_{m=1}^M$ are received by undersampling  channels $\{\mathbf{h}_m\}_{m=1}^M$ that share the same support. Hence, we develop the following distributed compressive sensing problem:
    \begin{equation}
  	\begin{aligned}
  		& \underset{\varphi,r}{\rm arg \ min} \sum_{m=1}^{M}\left\Vert \bm{\xi}_m\right\Vert_0 \\
  		& {\rm s.t.} \ \sum_m^M \left\Vert{\mathbf{y}}_m-\mathbf{W}^H_m\mathbf{B}\bm{\xi}_m\right\Vert_2^2\leq \epsilon,\\
  		&\ \ \ \  \ {\rm supp}(\bm{\xi}_1)=\cdots={\rm supp}(\bm{\xi}_M),
  	\end{aligned}
  	\tag{$\mathcal{P}_3$}
  \end{equation}
  where ${\rm supp}(\cdot)$ denotes the sparsity support.
  
To address this issue, we propose an OG-DOLS algorithm, which simultaneously  selects the column that minimizes all residual signals across various sensing matrices and progressively refines the on-grid parameters. 
In the context of compressive sensing, OLS is a greedy recovery algorithm with performance comparable to or better than orthogonal matching pursuit (OMP). The main distinction is in their column selection criteria: OLS minimizes the residual norm, while OMP prioritizes maximizing atom-residual correlation. They are equivalent when the sensing matrix is orthogonal.

Next, we start to introduce OG-DOLS from two main steps: atom identifying and off-grid refinement.

\begin{algorithm}
	\caption{Off-Grid Distributed Orthogonal Least Squares} 
	\label{AL1}
	\KwIn {Measured signals $\{\mathbf{y}_m\}_{m=1}^M$, random or optimized measurement matrices (Section \ref{MMO}) $\{\mathbf{W}_m\}_{m=1}^M$ , dictionary $\mathbf{B}$,  sparsity level $L$, $\widehat{\bm{\Phi}}^{(0)}_m=\emptyset$, and $\mathcal{L}$ representing the atom index set.
	}
	\KwOut {Parameters $\{\widehat{\varphi}_l,\widehat{r}_l\}_{l=1}^L$  and $\{\widehat{\bm{\xi}}_m\}_{m=1}^{M}$}
	\Begin{ 
		\For{$l=1,\cdots,L$}{
			 \%\%\%\%\%\%\% \emph{Atom Identifying} \%\%\%\%\%\%\% \\
			\For{$l\in\mathcal{L}$}{
				\For{$m=1,\cdots,M$}{
					$\widehat{\bm{\Phi}}^{(l)}_m=\left[\widehat{\bm{\Phi}}^{(l-1)}_m, \mathbf{W}_m^H[\mathbf{B}]_{:,l}\right]$.\\$\widehat{\bm{\Phi}}^{(l),\perp}_m= \mathbf{I}_P-\widehat{\bm{\Phi}}^{(l)}_m \widehat{\bm{\Phi}}^{(l),\dagger}_m$. \\
				$c_{l,m}=\left\Vert \widehat{\bm{\Phi}}^{(l),\perp}_m \mathbf{y}_m\right\Vert_2^2$.	}
			} 
			$ l^*=\underset{i\in\mathcal{I}}{\rm arg\ min} \ \sum_{m=1}^M c_{l,m}$.
			\\
			 $\mathcal{L}\leftarrow\mathcal{L}\setminus l^*$,  $\widehat{\bm{\Phi}}^{(l)}_m\leftarrow\left[ \widehat{\bm{\Phi}}^{(l-1)}_m, \mathbf{W}_m^H[\mathbf{B}]_{:,l^\star}\right]$. \\
			  $
			 \widehat{\bm{\xi}}_m= \widehat{\bm{\Phi}}^{(l),\dagger}_m\mathbf{y}_m$, $\forall m$.\\
			 Yield $\{\widehat{\varphi}_{k},\widehat{r}_{k}\}_{k=1}^l$.
			 \\
			 \%\%\%\%\%\% \emph{Off-Grid Refinement} \%\%\%\%\%\% \\
			 \For{$i=1,\cdots,I$}{
			 $\forall m:$ 
			 Calculate 
			$\mathbf{t}_m={\mathbf{y}}_m-\mathbf{W}^H_m\sum_{k=1}^{l}\widehat{\xi}_{m,k}\overline{\mathbf{b}}\left(\widehat{\varphi}_k,{\widehat{R}_k}\right)$  and
			construct $\mathbf{P}_m^\varphi$ and $\mathbf{P}_m^r$ according to Eqs. (\ref{PP}) and (\ref{PR}).\\
			 Obtain $\widetilde{\mathbf{P}}^{\varphi}$, $\widetilde{\mathbf{P}}^{r}$ and $\widetilde{\mathbf{t}}$.\\ $\bm{\eta}^\varphi=\Re\left\{\widetilde{\mathbf{P}}^{\varphi,\dagger}\widetilde{\mathbf{t}}\right\}$, $\bm{\eta}^r=\Re\left\{\widetilde{\mathbf{P}}^{r,\dagger}\widetilde{\mathbf{t}}\right\}$.\\
			 Update  $\widehat{\bm{\varphi}}\leftarrow \widehat{\bm{\varphi}}+ \bm{\eta}^\varphi\bullet \widehat{\bm{\varphi}}$ and $\widehat{\bm{r}}\leftarrow \widehat{\bm{r}}+ \bm{\eta}^r\bullet \widehat{\bm{r}}$.\\
			 Update $\widehat{\bm{\Phi}}^{(l),\dagger}_m$ and $
			 \widehat{\bm{\xi}}_m$ for $\forall m$.
		}}  
	}
		\Return{ $\{\widehat{\varphi}_l,\widehat{r}_l\}_{l=1}^L$  and $\{\widehat{\bm{\xi}}_m\}_{m=1}^{M}$.}
\end{algorithm}

\emph{Atom Identifying:} The core of DOLS is to find the current optimal index by miniziming the residual based on the established channel support $\widehat{\bm{\Phi}}^{(l-1)}_m=\mathbf{W}_m^H\left[\mathbf{b}(\widehat{\varphi}_1,\widehat{r}_1),\cdots,\mathbf{b}(\widehat{\varphi}_{l-1},\widehat{r}_{l-1})\right]$ in the $l-1$-th iteration. This suggests the $l$-th atom selection:
\begin{equation}\label{DOLS1}
	l^*=	\underset{l\in\mathcal{I}}{\rm arg\ min} \sum_{m=1}^M \left\Vert \mathbf{y}_m-\widehat{\bm{\Phi}}^{(l)}_m \widehat{\bm{\Phi}}^{(l),\dagger}_m \mathbf{y}_m\right\Vert_2^2,
\end{equation}
where $\widehat{\bm{\Phi}}^{(l)}_m=\left[\widehat{\bm{\Phi}}^{(l-1)}_m, \mathbf{W}_m^H[\mathbf{B}]_{:,l}\right]$.
For clarity, Eqn. (\ref{DOLS1}) is written as
\begin{equation}
	\begin{aligned}
		l^*&=\underset{i\in\mathcal{L}}{\rm arg\ min} \sum_{m=1}^M \left\Vert \mathbf{y}_m-\widehat{\bm{\Phi}}^{(l)}_m \widehat{\bm{\Phi}}^{(l),\dagger}_m \mathbf{y}_m\right\Vert_2^2 \\
		&=\underset{l\in\mathcal{L}}{\rm arg\ min} \ \sum_{m=1}^M \left\Vert \widehat{\bm{\Phi}}^{(l),\perp}_m \mathbf{y}_m\right\Vert_2^2,
	\end{aligned}
\end{equation}
where $\widehat{\bm{\Phi}}^{(l),\perp}_m\triangleq \mathbf{I}_P-\widehat{\bm{\Phi}}^{(l)}_m \widehat{\bm{\Phi}}^{(l),\dagger}_m$ denotes the projection matrix onto the span of the columns of $\widehat{\bm{\Phi}}^{(l)}_m$.

After $l^\star$ is determined, $\widehat{\bm{\Phi}}^{(l)}_m$ is updated by $\widehat{\bm{\Phi}}^{(l)}_m\leftarrow\left[ \widehat{\bm{\Phi}}^{(l-1)}_m, \mathbf{W}_m^H[\mathbf{B}]_{:,l^\star}\right]$.
 \emph{Off-Grid Refinement:}
Noticing $l^\star$ is selected as an on-grid point, it incurs a performance loss due to the off-grid parameter distribution.
To overcome this, the off-grid technique is adopted in each iteration to refine the estimated parameter. To initialize, we obtain the estimated angle and distance parameters from $\widehat{\bm{\Phi}}^{(l)}_m$, and derive the path gain
 $
\widehat{\bm{\xi}}_m= \widehat{\bm{\Phi}}^{(l),\dagger}_m\mathbf{y}_m$.  
Then, the parameter refinement in the $l$-th iteration problem is formulated by
    \begin{equation}\label{OG1}
	\begin{aligned}
		& \underset{\bm{\widehat{\xi}}_m,\widehat{\bm{\varphi}},\widehat{\bm{r}}}{\rm arg \ min}   \sum_m^M \left\Vert{\mathbf{y}}_m-\mathbf{W}^H_m\sum_{k=1}^{l}\widehat{\xi}_{m,k}{\mathbf{b}}\left(\widehat{\varphi}_k,{\widehat{r}_k}\right)
		\right\Vert_2^2. 
	\end{aligned}
	\tag{$\mathcal{P}_4$}
\end{equation}

Since the parameter $r$ is hard to directly deal with, we define $R\triangleq\frac{1}{r}$ and re-write the array manifold such that
$\overline{\mathbf{b}}(\varphi,R)\equiv \mathbf{b}(\varphi,r)$.
We use the permutation-based optimization method to solve problem (\ref{OG1}). Denoted by $\bm{\eta}^\varphi$ and $\bm{\eta}^r$ the permutations for $\varphi$ and $r$, respectively, the following problem is formulated:
     \begin{equation}\label{OG2}
	\begin{aligned}
		& \underset{\bm{\widehat{\xi}}_m,\bm{\eta}^\varphi,\bm{\eta}^r}{\rm arg \ min}   \sum_m^M \left\Vert{\mathbf{y}}_m-\mathbf{W}^H_m\sum_{k=1}^{l}\widehat{\xi}_{m,k}\overline{\mathbf{b}}(\widehat{\varphi}_k+\eta_k^\varphi \widehat{\varphi}_k,\widehat{R}_k+\eta_{k}^r  {\widehat{R}_k})
		\right\Vert_2^2 
	\end{aligned}
	\tag{$\mathcal{P}_5$}
\end{equation}

Then, using the first-order Taylor expansion for the permutated array manifold:
\begin{equation}
	\begin{aligned}
	&	\overline{\mathbf{b}}\left(\widehat{\varphi}_k+\eta_k^\varphi \widehat{\varphi}_k,\widehat{R}_k+\eta_{k}^r  {\widehat{R}_k}\right) \\ &\approx\overline{\mathbf{b}}(\widehat{\varphi}_k,\widehat{R}_k)+\eta_k^\varphi\widehat{\varphi}_k\frac{\partial \overline{\mathbf{b}}(\widehat{\varphi}_k,\widehat{R}_k)}{\partial \widehat{\varphi}_k}+ {\eta_k^r}{\widehat{R}_k}\frac{\partial \overline{\mathbf{b}}(\widehat{\varphi}_k,\widehat{R}_k)}{\partial  {\widehat{R}_k}},
	\end{aligned}
\end{equation}
where $\left[\frac{\partial \overline{\mathbf{b}}(\widehat{\varphi}_k,\widehat{R}_k)}{\partial \widehat{\varphi}_k}\right]_n=\frac{1}{\sqrt{N}}(j\frac{2\pi}{\lambda}(n-1)d-j\frac{2\pi}{\lambda}((n-1)d)^2\widehat{\varphi}_k\widehat{R}_k)e^{j\frac{2\pi}{\lambda}\left((n-1)d\widehat{\varphi}_k-\frac{((n-1)d)^2}{2}(1-\widehat{\varphi}_k^2)\widehat{R}_k\right)}$, and $\left[\frac{\partial \overline{\mathbf{b}}(\widehat{\varphi}_k,\widehat{R}_k)}{\partial \widehat{R}_k}\right]_n=-j\frac{1}{\sqrt{N}}\frac{2\pi}{\lambda} \frac{((n-1)d)^2}{2}(1-\widehat{\varphi}_k^2)
e^{j\frac{2\pi}{\lambda}\left((n-1)d\widehat{\varphi}_k-\frac{((n-1)d)^2}{2}(1-\widehat{\varphi}_k^2)\widehat{R}_k\right)}$.

The objective function in problem (\ref{OG2}) can also be written by
\begin{equation}
	\begin{aligned}
	&{\mathbf{y}}_m-\mathbf{W}^H_m\sum_{k=1}^{l}\widehat{\xi}_{m,k}\overline{\mathbf{b}}\left(\widehat{\varphi}_k+\eta_k^\varphi \widehat{\varphi}_k,\widehat{R}_k+\eta_{k}^r  {\widehat{R}_k}\right)=
	 \\
		&\mathbf{t}_m -\mathbf{W}^H_m\sum_{k=1}^{l}\widehat{\xi}_{m,k}\left(\eta_k^\varphi\widehat{\varphi}_k\frac{\partial \overline{\mathbf{b}}(\widehat{\varphi}_k,\widehat{R}_k)}{\partial \widehat{\varphi}_k}+{\eta_k^r}{\widehat{R}_k}\frac{\partial \overline{\mathbf{b}}(\widehat{\varphi}_k,\widehat{R}_k)}{\partial  {\widehat{R}_k}}
		\right),
	\end{aligned}
\end{equation}
where 
$\mathbf{t}_m\triangleq{\mathbf{y}}_m-\mathbf{W}^H_m\sum_{k=1}^{l}\widehat{\xi}_{m,k}\overline{\mathbf{b}}\left(\widehat{\varphi}_k,{\widehat{R}_k}\right)$.

By stacking the deriative vectors, we define
\begin{equation}\label{PP}
\mathbf{P}_m^\varphi\triangleq\left[\widehat{\xi}_{m,1}\widehat{\varphi}_1\frac{\partial \overline{\mathbf{b}}(\widehat{\varphi}_1,\widehat{R}_1)}{\partial \widehat{\varphi}_1},\cdots,\widehat{\xi}_{m,l}\widehat{\varphi}_l\frac{\partial \overline{\mathbf{b}}(\widehat{\varphi}_l,\widehat{R}_l)}{\partial \widehat{\varphi}_l}\right],
\end{equation}
 \begin{equation}\label{PR}
 \mathbf{P}_{m}^r\triangleq\left[\widehat{\xi}_{m,1}\widehat{R}_1\frac{\partial \overline{\mathbf{b}}(\widehat{\varphi}_1,\widehat{R}_1)}{\partial \widehat{R}_1},\cdots,\widehat{\xi}_{m,l}\widehat{R}_l\frac{\partial \overline{\mathbf{b}}(\widehat{\varphi}_l,\widehat{R}_l)}{\partial \widehat{R}_l}\right].
 \end{equation}

Fixing $\widehat{\bm{\xi}}$ and using the LS, the updating rule for $\bm{\eta}^\varphi$ and $\bm{\eta}^r$ is $
\bm{\eta}^\varphi=\left(\mathbf{W}^H_m\mathbf{P}^{\varphi}_m\right)^\dagger\mathbf{t}_m$ and $
\bm{\eta}^r=\left(\mathbf{W}^H_m\mathbf{P}^{r}_m\right)^\dagger\mathbf{t}_m$. Noticing that $\bm{\eta}^{\varphi/r}$ can be calculated for $\forall m$, we row-stack $\widetilde{\mathbf{P}}^{\varphi}\triangleq \left[\mathbf{P}^{\varphi,T}_1\mathbf{W}^*_1,\cdots,\mathbf{P}^{\varphi,T}_M\mathbf{W}^*_M\right]^T\in\mathbb{C}^{MP\times l}$, $\widetilde{\mathbf{P}}^{r}\triangleq \left[\mathbf{P}^{r,T}_1\mathbf{W}^*_1,\cdots,\mathbf{P}^{r,T}_M\mathbf{W}^*_M\right]^T\in\mathbb{C}^{MP\times l}$ and $\widetilde{\mathbf{t}}\triangleq[\mathbf{t}_1^T,\cdots,\mathbf{t}_M^T]^T\in\mathbb{C}^{MP\times 1}$. Thus to yield
\begin{equation} 
\bm{\eta}^\varphi=\Re\left\{\widetilde{\mathbf{P}}^{\varphi,\dagger}\widetilde{\mathbf{t}}\right\},
\end{equation}
\begin{equation} 
	\bm{\eta}^r=\Re\left\{\widetilde{\mathbf{P}}^{r,\dagger}\widetilde{\mathbf{t}}\right\},
\end{equation}
where the $\Re\{\cdot\}$ operator is used since $\varphi$ and $r$ should be real number. The estimated parameters are updated by $\widehat{\bm{\varphi}}\leftarrow \widehat{\bm{\varphi}}+ \bm{\eta}^\varphi\bullet \widehat{\bm{\varphi}}$ and $\widehat{\bm{r}}\leftarrow \widehat{\bm{r}}+ \bm{\eta}^r\bullet \widehat{\bm{r}}$.
Following that, the matrix $\{\widehat{\bm{\Phi}}^{(l)}_m\}_{m=1}^M$ is updated, and
the corresponding coefficient for $\forall m$ can be separately solved by the LS:  $
\widehat{\bm{\xi}}_m= \widehat{\bm{\Phi}}^{(l),\dagger}_m\mathbf{y}_m$.

  \subsubsection{Estimation of $\{\vartheta_l\}_{l=1}^L$}
According to Eqn. (\ref{gD}), $	\overline{\mathbf{h}}$ is approximated by
  \begin{equation}\label{Ha}
  	\overline{\mathbf{h}}\approx\sqrt{\frac{MN}{L}}\sum_{l=1}^{L}z_l\mathbf{a}(\vartheta_l)\otimes \mathbf{b}(\varphi_l,r_l).
  \end{equation}
  Then, we can obtain the approximated channel for the $m$-th microstrip:
  \begin{equation}
  	\begin{aligned}
  	\mathbf{h}_m=&\sqrt{\frac{MN}{L}}\sum_{l=1}^{L}z_l[\mathbf{a}(\vartheta_l)]_m \mathbf{b}(\varphi_l,r_l)\\
  	=&\widetilde{\mathbf{B}}\mathbf{z}_m,
  \end{aligned}
  \end{equation} 
  where $\mathbf{z}_m\triangleq\sqrt{\frac{MN}{L}}[z_1[\mathbf{a}(\vartheta_1)]_m,\cdots,z_L[\mathbf{a}(\vartheta_L)]_m]^T$, and $\widetilde{\mathbf{B}}\triangleq[\mathbf{b}(\varphi_1,r_1),\cdots,\mathbf{b}(\varphi_L,r_L)]\in\mathbb{C}^{N\times L}$.
  
  Denoted by $\mathbf{Z}\triangleq [\mathbf{z}_1,\cdots,\mathbf{z}_M]^T\in\mathbb{C}^{M\times L}$ and $\overline{\mathbf{z}}_l\triangleq [\mathbf{Z}]_{:,l}$, the estimation for $\widehat{\theta}_l, \forall l$, can be achieved by
  \begin{equation} \widehat{\vartheta}_l=\underset{\vartheta_l}{\rm arg \ max} \    {\left\vert  \mathbf{a}^H(\vartheta_l)\overline{\mathbf{z}}_l\right\vert}. 
  \end{equation}

Finally, $\{\vartheta_l,\varphi_l,r_l\}_{l=1}^L$ are obtained to calculate the channel gains and reconstruct the channel. 

  \section{Measurement Matrix Optimization}\label{MMO}
  Recalling the pilot signal model, the measurement vector $\mathbf{w}_{m,p}$ for the $m$-the RF chain in pilot $p$ is used to sample the channel. The quality of the measurement matrix  will significantly impact the channel estimation error. One simple measurement matrix is established by the complex Guassian random distribution. In this paper, MMO with the DMA constraint is discussed. 
  
 A widely adopted metric for sensing matrix optimization is formulated by minimizing the total coherence:
  \begin{equation}\label{IG1}
  	\begin{aligned}
  	&	\underset{\widetilde{\mathbf{W}}}{\rm arg \ min} \left\Vert \mathbf{I}_{\overline{g}}-\mathbf{B}^H\widetilde{\mathbf{W}}\widetilde{\mathbf{W}}^H\mathbf{B}\right\Vert_F^2 \\
  	&\ \ \ \ \ {\rm s.t.} \ [\mathbf{Q}]_{n,p}\in\mathcal{Q}, \ \forall n,p.
  	\end{aligned}
  \end{equation}
  With some derivations, the objective function is re-written as
  \begin{equation}
  	\begin{aligned}
  		&\left\Vert \mathbf{I}_{\overline{g}}-\mathbf{B}^H{\mathbf{W}}{\mathbf{W}}^H\mathbf{B}\right\Vert_F^2\\ &=\left\Vert \mathbf{I}_{\overline{g}}-\mathbf{B}^H{\mathbf{W}}{\mathbf{W}}^H\mathbf{B}\right\Vert_F^2 \\ 
  		&={\rm Tr}\left(\mathbf{I}_{\overline{g}}-2\mathbf{B}^H{\mathbf{W}}{\mathbf{W}}^H\mathbf{B}+\mathbf{B}^H{\mathbf{W}}{\mathbf{W}}^H\mathbf{B}\mathbf{B}^H{\mathbf{W}}{\mathbf{W}}^H\mathbf{B}\right)\\
  		&={\rm Tr}\left(\mathbf{I}_{P}-2{\mathbf{W}}^H\mathbf{B}\mathbf{B}^H{\mathbf{W}}+{\mathbf{W}}^H\mathbf{B}\mathbf{B}^H{\mathbf{W}}{\mathbf{W}}^H\mathbf{B}\mathbf{B}^H{\mathbf{W}}\right) \\
  		& \ \ \ \ \ + \overline{g}-P\\
  		&=\left\Vert \mathbf{I}_{P}-{\mathbf{W}}^H\mathbf{B}\mathbf{B}^H{\mathbf{W}}\right\Vert_F^2  + \overline{g}-P\\
  		&=\left\Vert \mathbf{I}_{P}-\mathbf{Q}^H{\mathbf{V}}\mathbf{B}\mathbf{B}^H\mathbf{V}^H\mathbf{Q}\right\Vert_F^2  + \overline{g}-P.
  	\end{aligned} 
  \end{equation}
In this sense, problem (\ref{IG1}) is transformed into
\begin{equation}\label{SMO0}
	\begin{aligned}
		&	\underset{{\mathbf{Q}}}{\rm arg \ min} \left\Vert \mathbf{I}_{P}-\mathbf{Q}^H{\mathbf{V}}\mathbf{B}\mathbf{B}^H\mathbf{V}^H\mathbf{Q}\right\Vert_F^2 \\
		&\ \ \ \ \ {\rm s.t.} \ [\mathbf{Q}]_{n,p}\in\mathcal{Q}, \ \forall n,p. 
	\end{aligned}
\tag{$\mathcal{P}_6$}
\end{equation}

To obtain one feasible solution, we decouple this problem into two subproblems (\ref{SMO1}) and (\ref{SMO2}), one is to solve $\widetilde{\bm{\Phi}}\triangleq\mathbf{Q}^H{\mathbf{V}}\mathbf{B}$ without the LC, and another is to solve $\mathbf{Q}$ from $\widetilde{\bm{\Phi}}$ with the LC. Therefore, we have
  \begin{equation}\label{SMO1}
  	\begin{aligned}	\underset{\widetilde{\bm{\Phi}}}{\rm arg \ min} \left\Vert\mathbf{I}_{P}- \widetilde{\bm{\Phi}}\widetilde{\bm{\Phi}}^H\right\Vert_F^2. 
  	\end{aligned}
  \tag{$\mathcal{P}_{6.1}$}
  \end{equation}
The solution of this problem is discussed in Appendix \ref{appendixB}, following a expression of
\begin{equation}
\widetilde{\bm{\Phi}}=\frac{\overline{\sigma}^2}{P}\mathbf{{U}}_1[\mathbf{I}_{P}, \mathbf{0}_{{P},\overline{g}-{P}}]^T\mathbf{{U}}_2^H,
\end{equation}
 where $\overline{\sigma}^2$ denotes the power of $\mathbf{Q}^H\mathbf{V}\mathbf{B}$, and $\mathbf{{U}}_1\in\mathbb{C}^{P\times P}$ and $\mathbf{{U}}_2\in\mathbb{C}^{\overline{g}\times\overline{g}}$ are arbitrary unitary matrices.  

Then, the second subproblem solving $\mathbf{Q}$ is established by
\begin{equation}\label{SMO2}
	\begin{aligned}
		&	\underset{{\mathbf{Q}}}{\rm arg \ min} \left\Vert\widetilde{\bm{\Phi}}-{\mathbf{Q}}^H\mathbf{VB}\right\Vert_F^2  \\
		&\ \ \ \ \ {\rm s.t.} \ [\mathbf{Q}]_{n,p}\in\mathcal{Q}, \ \forall n,p.
	\end{aligned}
\tag{$\mathcal{P}_{6.2}$}
\end{equation}

Since ${\mathbf{Q}}\triangleq\frac{\mathbf{J}+\mathbf{F}}{2}$, with $\mathbf{J}$ representing a full $j$ matrix, we have the following simplification:
\begin{equation}
	\begin{aligned}
		\left\Vert\widetilde{\bm{\Phi}}-{\mathbf{Q}}^H\mathbf{VB}\right\Vert_F^2=& 	\left\Vert\widetilde{\bm{\Phi}}^H-\frac{1}{2}\mathbf{J}^H\mathbf{VB}-\frac{1}{2} \mathbf{F}^H\mathbf{VB}\right\Vert_F^2
		\\
		=&{\rm Tr} \left\{{\bm{\Psi}}^H{\bm{\Psi}}-\Re\left\{\mathbf{B}^H\mathbf{V}^H\mathbf{F}\bm{\Psi}\right\} \right. \\ & \left. \ \  \ \  \ + \frac{1}{4}\mathbf{B}^H\mathbf{V}^H {\mathbf{F}}
	{\mathbf{F}}^H\mathbf{V}\mathbf{B}	\right\},
	\end{aligned}
\end{equation}
where $\bm{\Psi}\triangleq\widetilde{\bm{\Phi}}^H-\frac{1}{2} \mathbf{J}^H\mathbf{VB}$. Thus, the second optimization problem is written as
\begin{equation} 
	\begin{aligned}
		&	\underset{{\mathbf{F}}}{\rm arg \ min} \
		\frac{1}{4}{\rm Tr}\left\{{\mathbf{F}}{\mathbf{F}}^H\mathbf{V}\mathbf{B}\mathbf{B}^H\mathbf{V}^H\right\} -{\rm Tr}\left\{\Re\left\{\mathbf{F}\bm{\Psi}\mathbf{B}^H\mathbf{V}^H\right\}\right\} \\
		&\ \ \ \ \ \ \  \ \ \ \ \ {\rm s.t.} \ [\mathbf{F}]_{n,p}\in\mathcal{F}, \ \forall n,p,
	\end{aligned}
	\tag{$\mathcal{P}_{7}$}
\end{equation}
where $\mathcal{F}\triangleq\{e^{jx}|x\in[0,2\pi]\}$.

For clarity of the following derivation, we denote by $\overline{\mathbf{F}}\triangleq\mathbf{F}^H$, $\mathbf{X}_1\triangleq \frac{1}{4} \mathbf{V}\mathbf{B}\mathbf{B}^H\mathbf{V}^H$, and $\mathbf{X}_2\triangleq \frac{1}{2}\bm{\Psi}\mathbf{B}^H\mathbf{V}^H$, and formulate
\begin{equation} 
	\begin{aligned}
		&	\underset{\overline{\mathbf{F}}}{\rm arg \ min} \
	 {\rm Tr}\left\{\overline{\mathbf{F}}^H\overline{\mathbf{F}}\mathbf{X}_1\right\} -{\rm Tr}\left\{2\Re\left\{\overline{\mathbf{F}}^H\mathbf{X}_2\right\}\right\} \\
		&\ \ \ \ \ \ \ \ {\rm s.t.} \ \left[\overline{\mathbf{F}}\right]_{p,n}\in\mathcal{F}, \ \forall n,p,
	\end{aligned}
	\tag{$\mathcal{P}_{8}$}
\end{equation}
This quadratic problem is addressed as follows.
Denoted by the objective function
$f(\overline{\mathbf{F}})\triangleq 	 {\rm Tr}\left\{\overline{\mathbf{F}}^H\overline{\mathbf{F}}\mathbf{X}_1\right\} -{\rm Tr}\left\{2\Re\left\{\overline{\mathbf{F}}^H\mathbf{X}_2\right\}\right\}$, its $(p,n)$-th entry is expressed by 
\begin{equation}
f([\overline{\mathbf{F}}]_{p,n})={\kappa_{n,n}}\vert[\overline{\mathbf{F}}]_{p,n}\vert^2-2\Re\left\{\nu^*_{p,n} [\overline{\mathbf{F}}]_{p,n}\right\}.
\end{equation}

Noticing that $\vert[\overline{\mathbf{F}}]_{n,p}\vert^2=1$, by minimizing $-\Re\left\{\nu_{p,n}^* [\overline{\mathbf{F}}]_{p,n}\right\}$, the element-wise closed-form solution is given by
\begin{equation}
\left[\overline{\mathbf{F}}\right]_{p,n}=\frac{\nu_{p,n}}{\vert\nu_{p,n}\vert}.
\end{equation}
Then, we derive $\kappa_{n,n}$ and $\nu_{n,p}$. Considering the derivative of the objective function with respect to $\overline{\mathbf{F}}^*$:
\begin{equation}
	\begin{aligned}
		\frac{\partial f(\overline{\mathbf{F}})}{\partial \overline{\mathbf{F}}^*}
		=&\overline{\mathbf{F}} \mathbf{X}_1-\mathbf{X}_2.
	\end{aligned}
\end{equation}
We can obtain its $(p,n)$-th entry as
\begin{equation}\label{com1}
	\begin{aligned}
		\left[	\frac{\partial f(\overline{\mathbf{F}})}{\partial \overline{\mathbf{F}}^*}\right]_{p,n}=&\sum_{\bar{n}}[\mathbf{X}_1]_{\bar{n},n}[\overline{\mathbf{F}}]_{p,\bar{n}}-[\mathbf{X}_2]_{p,n}\\
		=& [\mathbf{X}_1]_{n,n}[\overline{\mathbf{F}}]_{p,n}+ \sum_{\bar{n}\neq n}[\mathbf{X}_1]_{\bar{n},n}[\overline{\mathbf{F}}]_{p,\bar{n}}-[\mathbf{X}_2]_{p,n}.
	\end{aligned}
\end{equation}
Moreover, we can derive its equivalent expression as
\begin{equation}\label{com2}
	\frac{\partial f\left(\left[\overline{\mathbf{F}}\right]_{p,n}\right)}{\partial \left[\overline{\mathbf{F}}\right]_{p,n}^*}= {\kappa_{n,n}}\left[\overline{\mathbf{F}}\right]_{p,n}-{\nu_{p,n}}.
\end{equation}

Comparing Eqs. (\ref{com1}) and (\ref{com2}), we can yield $\kappa_{n,n}\equiv [\mathbf{X}_1]_{n,n}$ and $\nu_{p,n}\equiv [\mathbf{X}_2]_{p,n}-  \sum_{\bar{n}\neq n}[\mathbf{X}_1]_{\bar{n},n}[\overline{\mathbf{F}}]_{p,\bar{n}}$.

The above procedure applies for any dictionaries for MMO. However, noticing that the dictionary used in this paper satisfies $\mathbf{B}\mathbf{B}^H=\frac{\overline{g}}{N}\mathbf{I}_{N}$, thus combining problems (\ref{SMO0}) and (\ref{SMO2}) yields a simpler version:
\begin{equation}\label{SMO3}
	\begin{aligned}
		&	\underset{{\mathbf{F}}}{\rm arg \ min} \left\Vert\widetilde{\bm{\Phi}}-\frac{\mathbf{J}^H}{2}-\frac{\mathbf{F}^H}{2}\right\Vert_F^2  \\
		&\ \ \ \ \  {\rm s.t.} \ [\mathbf{F}]_{n,p}\in\mathcal{F}, \ \forall n,p.
	\end{aligned}
	\tag{$\mathcal{P}_{9}$}
\end{equation}

This problem can be solved with phase alignment such that $[\mathbf{F}]_{n,p}=e^{j\angle [\overline{\bm{\Psi}}]_{n,p}}$ with $\overline{\bm{\Psi}}\triangleq 2\widetilde{\bm{\Phi}}^H-\mathbf{J}$.
\begin{figure} 
	\centering
	\subfigure[A UPA of $128\times 8$]{
		\includegraphics[width=3.33in]{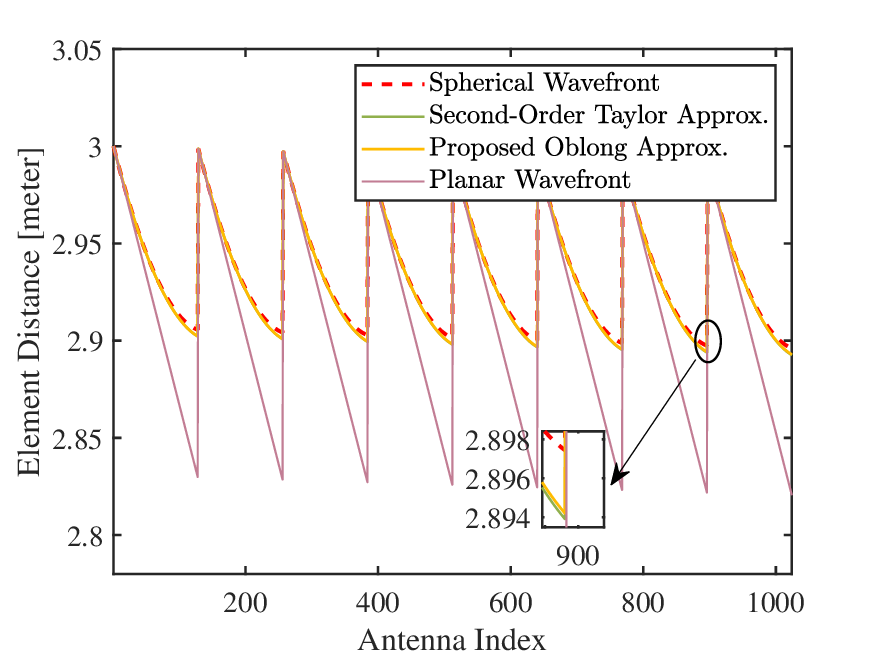}
	}
	\quad    
	\subfigure[A UPA of $128\times 32$]{
		\includegraphics[width=3.33in]{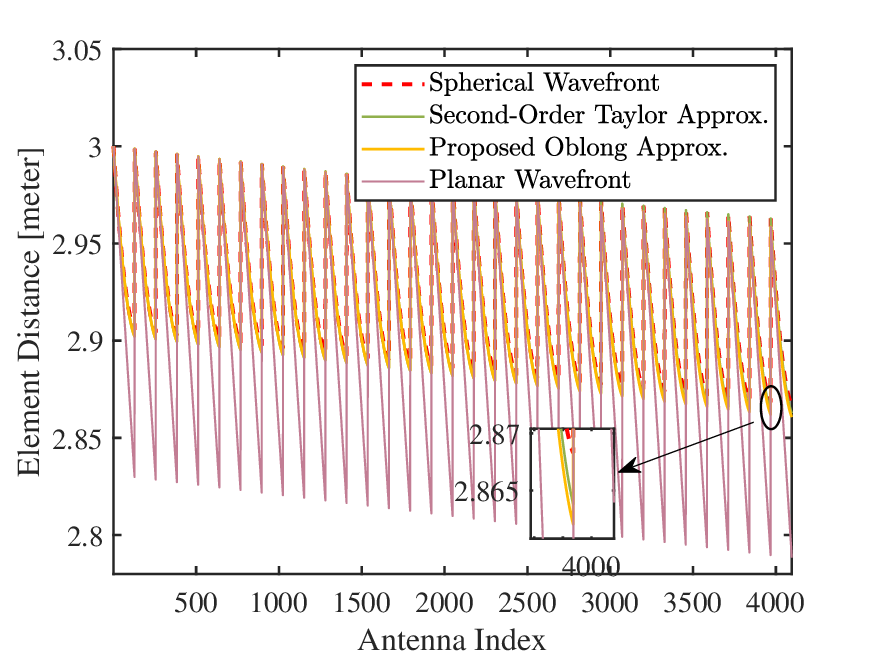}
	}
	\caption{Element distance versus the antenna index with different models. $r=3$, $\vartheta=\varphi=0.25$, and $\lambda=0.0107$.}
	\label{ED}
\end{figure}
\begin{figure} 
	\centering
	\subfigure[$\vartheta=\varphi=0.25$]{
		\includegraphics[width=3.33in]{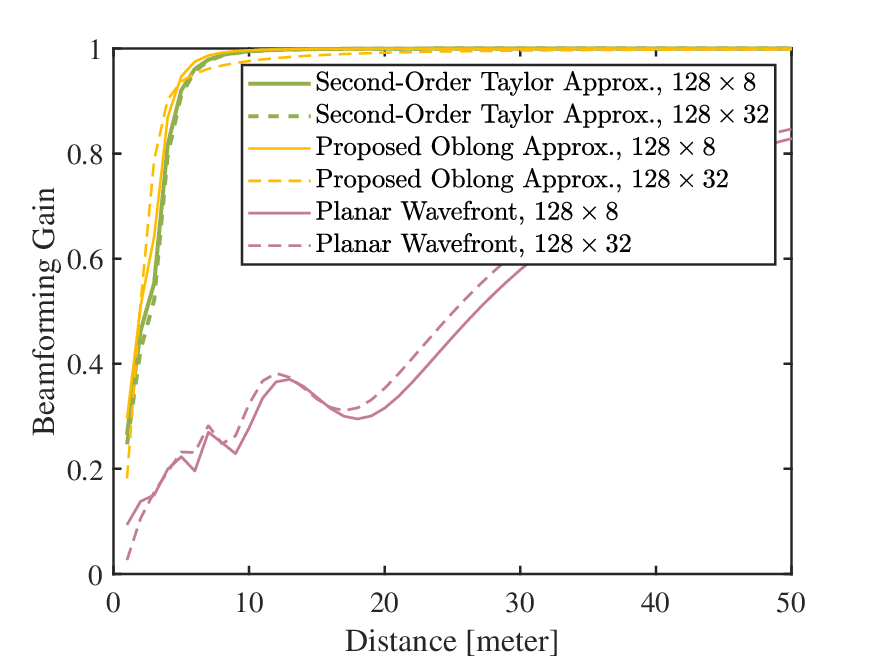}
	}
	\quad    
	\subfigure[$\vartheta=\varphi=0.5$]{
		\includegraphics[width=3.33in]{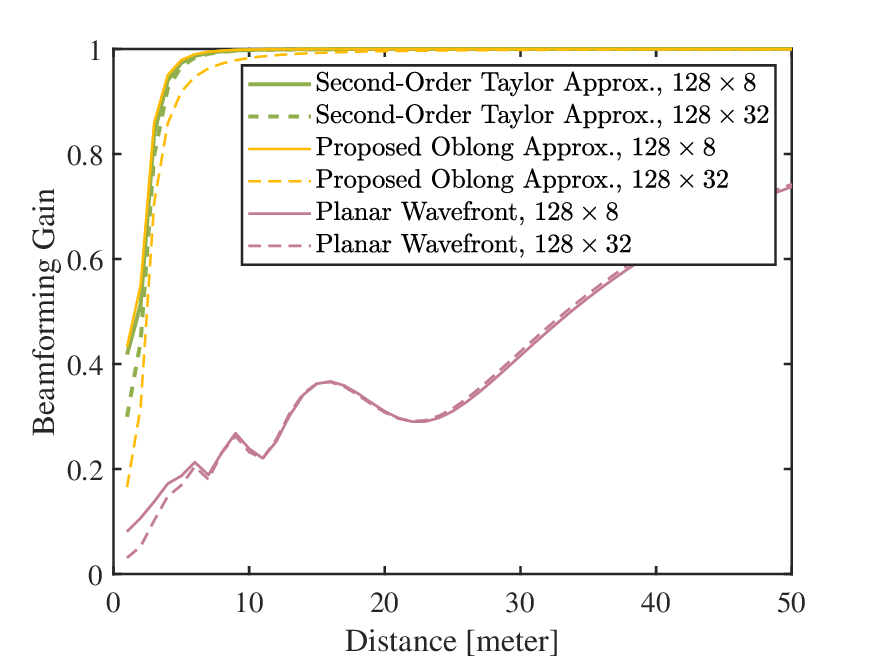}
	}
	\caption{Beamforming gain versus the communication distance with $\lambda=0.0107$.}
	\label{BF}
\end{figure} 
\section{Simulation Results}\label{simu}
In this section, our objective is to validate the effectiveness of the proposed methods in this paper. This involves assessing the error of the proposed Oblong Approx. model, evaluating the estimation performance with and without MMO, examining the channel estimation performance and parameter sensing performance using the proposed algorithms, and conducting a comparison between DMAs and phased-arrays regarding estimation error. The system parameters are set as follows. The system adopts $28$ GHz central frequency. Each microstrip consists of $N=128$ elements, and the number of RF chains varies in different simulations. The inter-antenna spacing is set to $d=\lambda/2$. The user/scatter locations are assumed to be distributed with $\vartheta,\varphi\in[-1,1]$ and $r\in[5,100]$ with channel paths $L=3$. Since the transmit power is set to 1, the signal-to-noise ratio (SNR) is defined by $\frac{1}{\sigma_n^2}$. The number of samples for dictionaries $\overline{\mathbf{G}}$ and $\mathbf{B}$ are set to $\overline{G}=2N*R*2M$ and $\overline{g}=2N$, respectively, where $2N$ is to sample the AZ angle, $R=20$ is to sample the AZ distance, and $2M$ is to sample the EL angle.
The main methods involved in these evaluations include as follows. Note that these methods are based on MMO and under the LC. 
\begin{itemize}
\item \textbf{EL-AZ-JE:} Solving problem (\ref{DE}) in Section \ref{ELAZJE} with OLS.
\item \textbf{AZ-IE:} Solving problem (\ref{IE}) in \ref{AZIE} with OLS.
\item \textbf{EL-AZ-DE:} Solving the two-stage recovery problem in Section \ref{ELAZDE}, in which the first stage uses the DOLS algorithm.
\item \textbf{OG-EL-AZ-DE:} Solving the two-stage recovery problem in Section \ref{ELAZDE}, in which the first stage uses the OG-DOLS algorithm.
\item \textbf{Oracle LS:} Assuming perfect channel support  to estimate the channel path gain and reconstruct the channel, which serves the lower bound in this study.
\end{itemize}
Moreover, we use $\textbf{w/o}$ $\text{LC}$ and $\textbf{w/o}$ $\text{MMO}$ to denote the method with the ideal PA and with a random Gaussian measurement matrix, respectively.

\begin{figure} 
	\centering
	\subfigure[$M=4$, $P=20$]{
		\includegraphics[width=3.45in]{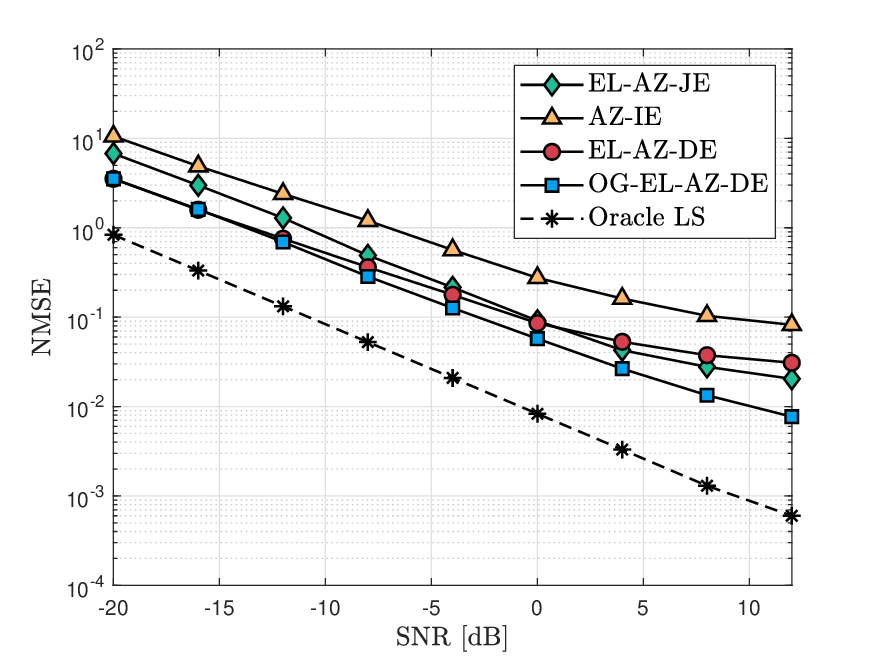}
	}
	\quad    
	\subfigure[$M=4$, $P=10$]{
		\includegraphics[width=3.45in]{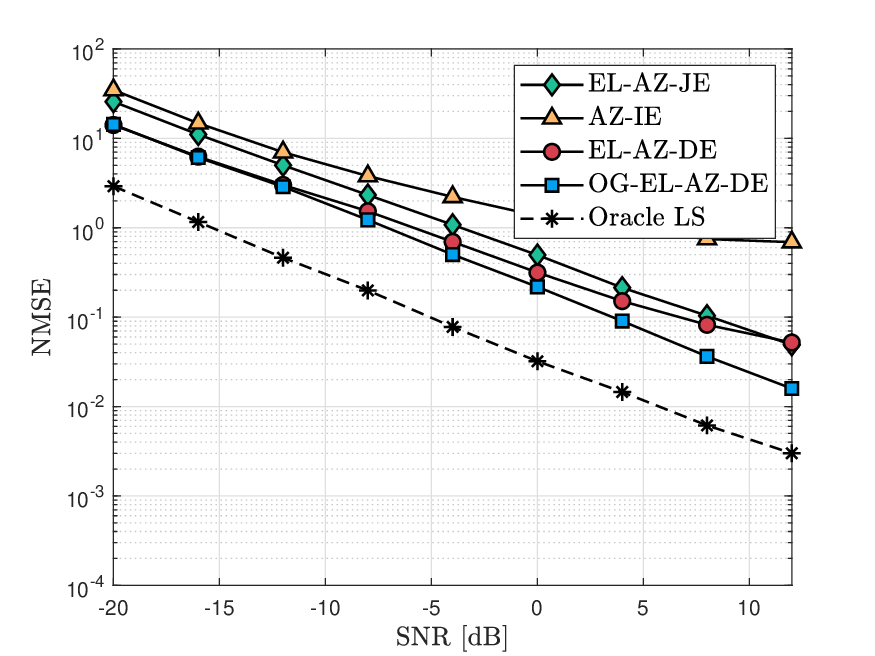}
	}
	\quad    
	\subfigure[$M=8$, $P=20$]{
		\includegraphics[width=3.45in]{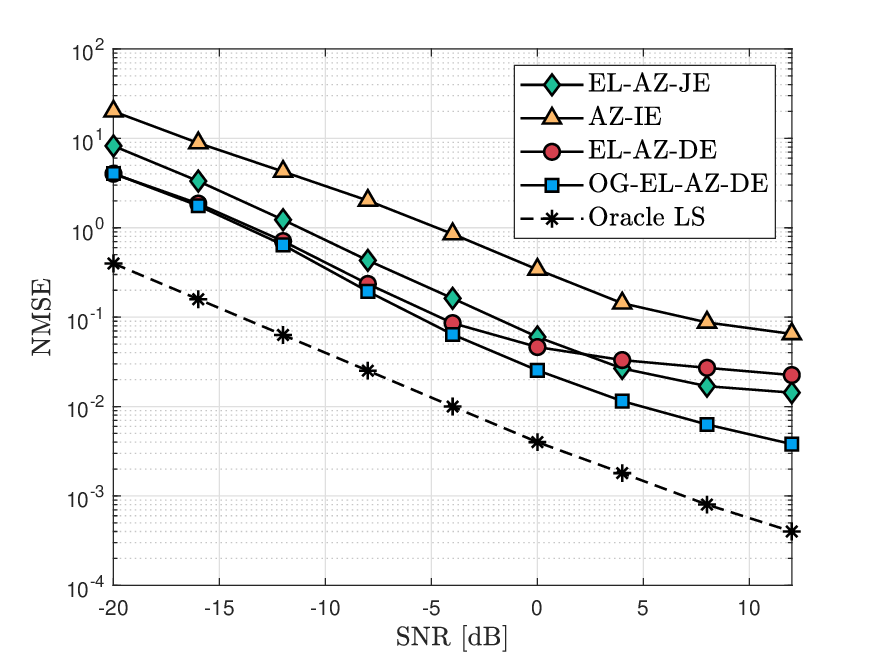}
	}
	\caption{The NMSE versus SNR of different methods. MMO for DMAs is considered.}
	\label{MDMA}
\end{figure}

We first conduct calculations and generate distance plots for each antenna element to the object (element distance) using various models. In these calculations, we assumed a reference antenna-to-object distance $r$ of 3 meters, with $\vartheta=\varphi=0.25$, and an antenna wavelength of 0.0107 meters. As depicted in Fig. \ref{ED}(a), we employed a $128\times8$ UPA, and the curves showing the element distances relative to the antenna index (stacked along the $y$-axis) have 8 periods. It is noticeable that under the planar-wave model, there is a significant discrepancy between the element distances and those under the spherical-wave model within each period. This difference primarily arises from the pronounced near-field effects attributed to the increased $y$-axis aperture. Conversely, it is evident that both Second-order Taylor Approx. and our proposed Oblong Approx. align more closely with the spherical-wave model.
Subsequently, in Fig. \ref{ED}(b), we extended the array size of $128\times 8$ to $128\times 32$, which accentuates the gap between Second-order Taylor Approx. and Oblong Approx. compared to Fig. \ref{ED}(a), yet it remains within an acceptable range.

As illustrated in Fig. \ref{BF}(a), where $\vartheta=\varphi=0.25$, it is observed that as the communication distance increases to approximately $5$ meters, the beamforming gain of the Oblong Approx. and Second-Order Taylor Approx. approaches to $1$. This signifies their capability to closely match the spherical-wave model as the distance increases. In contrast, it is challenging for the planar wavefront to approximate the spherical-wave model in the near-field region, especially at distances less than $50$ meters. In Fig. \ref{BF}(b), with $\vartheta=\varphi=0.5$, increasing the number of EL elements from $8$ to $32$ results in only a minor degradation in the performance of Oblong Approx. This indicates the effectiveness of the Oblong Approx. model when applied for near-field oblong-shape arrays.

In the subsequent simulations, we utilize the normalized mean square error (NMSE) to assess the channel estimation performance of various estimation methods. Fig. \ref{MDMA} depicts the NMSE of different methods across a SNR range from $-20$ to $12$ dB, with $M\in\{4,8\}$ and $P\in\{10,20\}$.
Observing Fig. \ref{MDMA}(a), it is evident that, excluding the lower bound benchmark, the OG-EL-AZ-DE method outperforms all others in all SNR settings. This implies that the decoupled EL-AZ estimation framework can achieve robust estimation performance. A comparison between OG-EL-AZ-DE and EL-AZ-DE reveals the crucial role played by the proposed OG procedure, particularly as the SNR increases. Without the OG procedure, EL-AZ-DE exhibits a convergent trend with rising SNR, as the atom mismatch problem causes increased performance degradation, which the OG procedure mitigates to some extent.
Conversely, EL-AZ-JE, being a global greedy method, demonstrates poorer performance than EL-AZ-DE at low SNR. This is attributed to the large-scale dictionary inherent in EL-AZ-JE, leading to search redundancy and a negative impact on atom selection under unfavorable SNR conditions. Additionally, AZ-IE, with its simplest implementation, exhibits poor NMSE performance due to its failure to fully exploit the spatial near-field characteristics.

In Figs. \ref{MDMA}(b) and (c), where $P=10$ and $M=8$ are respectively set, a similar NMSE changing trend is observed compared to Fig. \ref{MDMA}(a). As the number of measurements decreases to $P=10$, the estimation performance diminishes due to the smaller number of measurements. Conversely, setting the number of RF chains to $M=8$ results in a slight improvement in estimation performance due to enhanced array gain.

  \begin{figure}
	\centering
	\includegraphics[width = 0.496\textwidth]{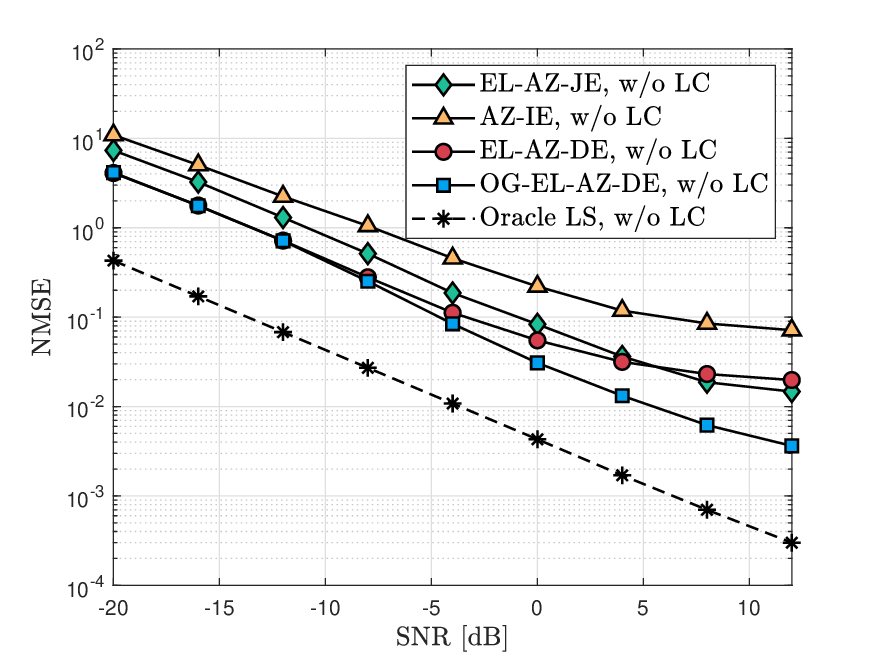}
	\caption{The NMSE versus SNR of different methods. $M=4$, $P=20$. MMO for phased-arrays is considered. }
	\label{MPA_M4_P20}
\end{figure} 
Next, we explore the impact of the LC of DMAs on NMSE performances. To this end, we consider a phased-array with  $q\in\{e^{jx},x\in[0,2\pi]\}$ to design the MMO process for channel estimation, as shown in Fig. \ref{MPA_M4_P20}, where $M=4$, $P=20$, and SNR ranges from $-20$ to $12$ dB. Comparing Figs. \ref{MDMA}(a) and \ref{MPA_M4_P20}, we can conclude that under the MMO process, the LC significantly diminishes the estimation performance. For example, OG-EL-AZ-DE in Fig. \ref{MPA_M4_P20} achieves an NMSE of about $4e^{-3}$ at SNR = $12$ dB, whereas this value changes to about $8e^{-3}$ at the same setting with DMAs in Fig. \ref{MDMA}(a). This performance degradation is also evident when comparing other methods. The fundamental reason for this is that the LC restricts the feasible phase to $[0,\pi]$, as illustrated in Fig. \ref{DMAPA}. 

  \begin{figure}
	\centering
	\includegraphics[width = 0.496\textwidth]{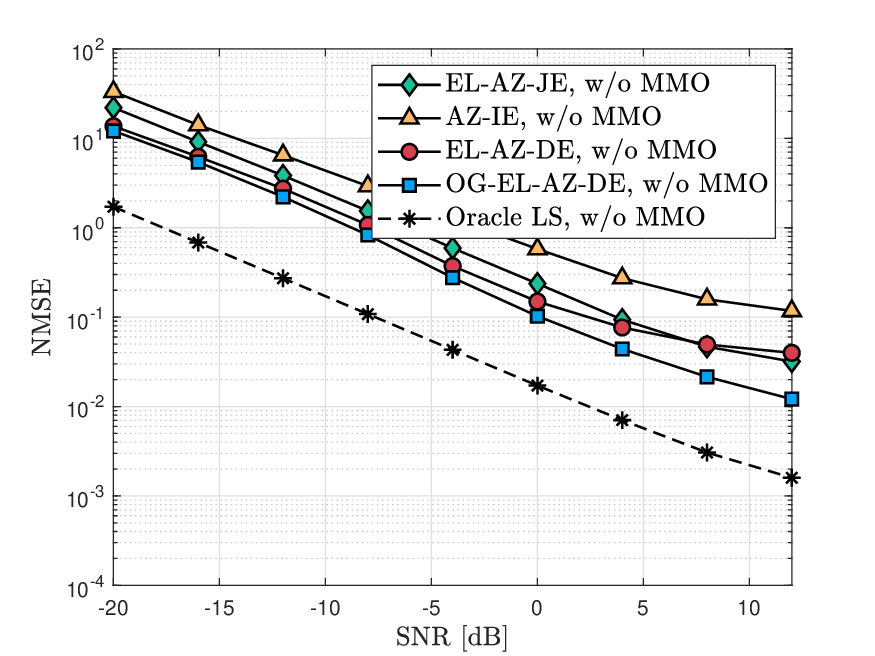}
	\caption{The NMSE versus SNR of different methods. $M=4$, $P=20$. Gaussian random measurement matrix for DMAs is considered. }
	\label{NOMDMA_M4_P20}
\end{figure}

\begin{figure}
  	\centering
  	\includegraphics[width = 0.496\textwidth]{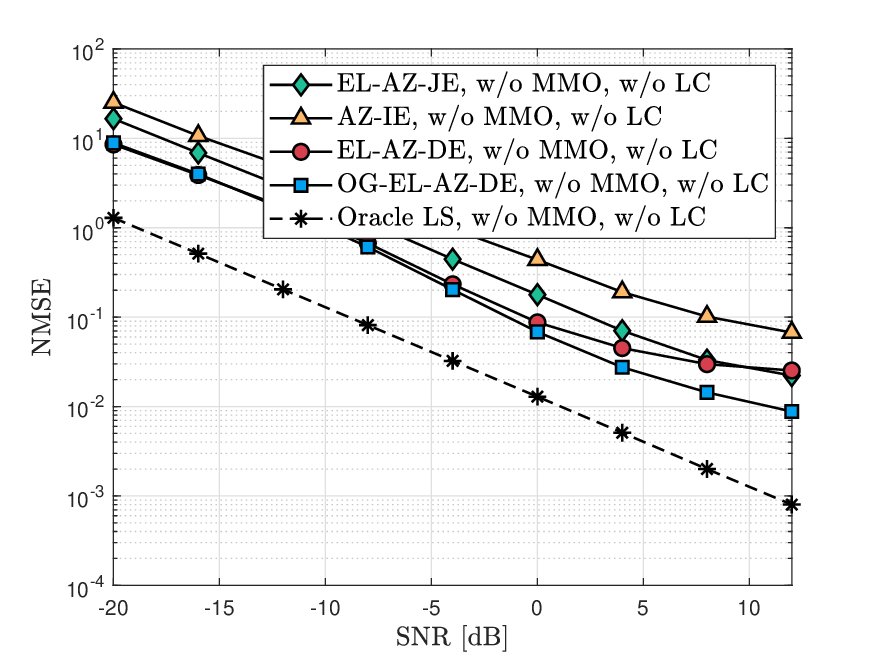}
  	\caption{The NMSE versus SNR of different methods. $M=4$, $P=20$. Gaussian random measurement matrix for phased-arrays is considered.}
  	\label{NOMPA_M4_P20}
  \end{figure}

To evaluate the impact of MMO on the NMSE performance of different methods, we present the results in Fig. \ref{NOMDMA_M4_P20}, where the setup is identical to that in Fig. \ref{MDMA}(a) except for the use of a Gaussian random measurement matrix instead of an optimized measurement matrix. By comparing Figs. \ref{MDMA}(a) and \ref{NOMDMA_M4_P20}, we observe an improvement in estimation performance with the adoption of MMO.  Additionally, Fig. \ref{NOMPA_M4_P20} replicates this simulation on phased-arrays, revealing that NMSE performances for phased-arrays without MMO are also worse than those with MMO in Fig. \ref{NOMDMA_M4_P20}.

  \begin{figure}
	\centering
	\includegraphics[width = 0.496\textwidth]{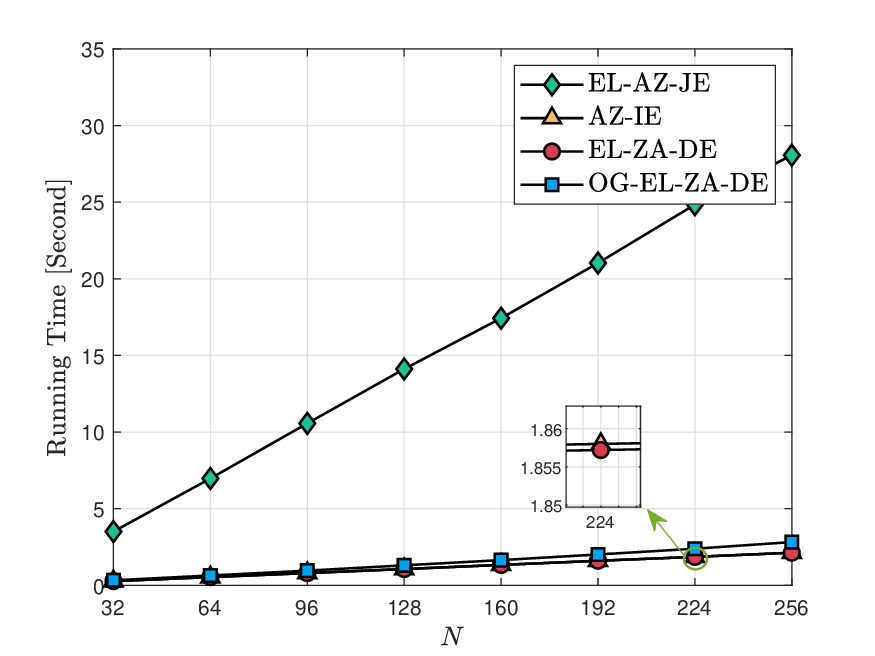}
	\caption{The running time versus $N$.}
	\label{Time}
\end{figure}  
 
 Finally, Fig. \ref{Time} presents the running time of different methods, with $M=4$, $P=20$, and $N$ ranging from $32$ to $256$. It is noteworthy that all methods exhibit linear time complexity with the number of metasurface elements $N$. AZ-IE and EL-AZ-DE demonstrate comparable speeds, while OG-EL-AZ-DE is slightly slower due to the additional OG procedure. In contrast, EL-AZ-JE has the slowest speed as it necessitates processing a large-scale angle-distance dictionary.
 \vspace{-0.2cm}
\section{Conclusions}\label{Con}
This paper concentrates on near-field channel estimation for DMAs, an emerging transceiver array architecture for XL-MIMO, recognized for its low hardware costs and energy efficiency. Initially, we establish a near-field propagation model for oblong-shaped arrays to decouple the EL-AZ parameters, referred to as the Oblong Approx., which can seamlessly be applied to XL-DMAs. The numerical evaluation indicates that the Oblong Approx model closely aligns with the second-order Taylor model, signifying its success as a model simplification. Subsequently, we propose four channel estimation frameworks---EL-AZ-JE, AZ-IE, EL-AZ-DE, and OG-EL-AZ-DE---from the perspective of jointly recovering EL-AZ angles and distance $\{\vartheta,\varphi,r\}$, individually recovering $\{\varphi_m,r_m\}_{m=1}^M$ for each microstrip, and recovering $\{\vartheta,\varphi,r\}$ in a two-stage procedure with/without off-grid parameter refinement. Notably, EL-AZ-DE and OG-EL-AZ-DE leverage the Oblong Approx model to implement a decoupled two-stage EL-AZ estimation, showcasing commendable NMSE performance and low linear complexity. Furthermore, we demonstrate the degradation in estimation performance of DMAs compared with ideal phased arrays, attributable to the impact of LC. Overall, the parameter estimation loss can be effectively addressed by enlarging the array aperture, given its cost-effectiveness.

The findings of this paper can be further extended. We have outlined several potential avenues for future research:
\begin{itemize}
	\item \emph{Better Recovery Algorithms:} 
Observing the simulation results, it is evident that the top-performing method, OG-EL-ZA-DE, still exhibits approximately a $10$ dB NMSE performance gap compared to the lower bound. Therefore, exploring advanced recovery algorithms, such as those based on Bayesian learning, may yield improved performance. 
	\item \emph{MMO for DMAs:} This paper designs MMO for DMAs with a focus on total coherence minimization and addresses the LC in a straightforward manner. Employing more efficient strategies could enhance this aspect further.
	
	\item \emph{Other UPA-based Applications:} Our proposed Oblong Approx. model is not only applicable to XL-DMAs but is also general for oblong-shaped arrays. This concept can be extended to intelligent reflecting surfaces, especially when they assume a low spatial resolution in the EL direction. It is worth noting that certain communication systems are ground-based, necessitating lower EL resolution.
	
\end{itemize}
\begin{appendices}
\section{  }\label{appendixB}
Using the SVD of $\widetilde{\bm{\Phi}}=\overline{\mathbf{U}}_1{\bm{\Sigma}}\overline{\mathbf{U}}_2^H$ with unitary matrices $\overline{\mathbf{U}}_1\in\mathbb{C}^{P\times P}$, $\overline{\mathbf{U}}_2\in\mathbb{C}^{\overline{g}\times \overline{g}}$ and $\bm{\Sigma}=\left[{\rm diag}(\bm{\sigma}), \mathbf{0}_{P,\overline{g}-P}\right]^T$, where $\bm{\sigma}=[\sigma_1,\cdots,\sigma_{P}]$ and $\mathbf{0}_{P,\overline{g}-P}\in\mathbb{C}^{P\times(\overline{g}-P)}$ is a null matrix. 
\begin{equation}
	\begin{aligned}
		\left\Vert\mathbf{I}_{P}-\widetilde{\bm{\Phi}}\widetilde{\bm{\Phi}}^H
		\right\Vert_F^2=&\left\Vert \mathbf{I}_{P}-\overline{\mathbf{U}}_1{\bm{\Sigma}}\overline{\mathbf{U}}_2^H\overline{\mathbf{U}}_2{\bm{\Sigma}}^H\overline{\mathbf{U}}_1^H
		\right\Vert_F^2\\
		=&\left\Vert \overline{\mathbf{U}}_1 \left(\mathbf{I}_{P}-{\bm{\Sigma}}{\bm{\Sigma}}^H\right)\overline{\mathbf{U}}_1^H
		\right\Vert_F^2\\
		=&\left\Vert
		\mathbf{I}_{P}-{\bm{\Sigma}}{\bm{\Sigma}}^H
		\right\Vert_F^2\\
		=&\sum_{i=1}^{P}(1-\sigma^2_i)^2.
	\end{aligned}
\end{equation}
We assume that $\left\Vert\widetilde{\bm{\Phi}}\right\Vert_F^2=\overline{P}$. Furthermore, since ${\rm tr}({\bm{\Phi}}{\bm{\Phi}}^H)={\rm tr}({\bm{\Sigma}}{\bm{\Sigma}}^H)=\sum_{i=1}^{P}\sigma^2_i$, the optimization problem w.r.t. $\widetilde{\bm{\Phi}}$ is expressed as
\begin{equation}
	\begin{aligned}
		&\underset{\bm{\sigma}^2}{\rm arg \ min} \ \sum_{i=1}^{P}(1-\sigma^2_i)^2 \\
		&{\rm s.t.} \ \sum_{i=1}^{P}\sigma^2_i=\overline{\sigma}^2.
	\end{aligned}
\end{equation}
This is a convex problem w.r.t. $\bm{\sigma}^2$, and can be solved by the method of Lagrange multipliers. Considering the Lagrange function $\mathcal{L}(\bm{\sigma}^2,\lambda_\mathcal{L})=\sum_{i=1}^{P}(1-\sigma^2_i)^2+\lambda_\mathcal{L}(P-\sum_{i=1}^{P}\sigma^2_i)$ with $\lambda_\mathcal{L}$ denoting the Lagrange multiplier. The solution can be derived by setting the partial derivatives w.r.t. $\sigma^2$ and $\lambda_\mathcal{L}$ to zero, i.e., $\frac{\partial \mathcal{L}(\bm{\sigma}^2,\lambda_\mathcal{L})}{\partial \bm{\sigma}^2}=\frac{\partial \mathcal{L}(\bm{\sigma}^2,\lambda_\mathcal{L})}{\partial \lambda_\mathcal{L}}=0$. Then we can obtain its solution $\sigma^2_{1}=\cdots=\sigma^2_{P}=\frac{\overline{\sigma}^2}{P}$. Hence, $\widetilde{\bm{\Phi}}=\frac{\overline{\sigma}^2}{P}{\mathbf{{U}}}_1[\mathbf{I}_{P}, \mathbf{0}_{P,\overline{g}-P}]^T{\mathbf{{U}}}_2^H$, where ${\mathbf{{U}}}_1\in\mathbb{C}^{\overline{g}\times\overline{g}}$ and ${\mathbf{{U}}}_2\in\mathbb{C}^{P\times P}$ are arbitrary unitary matrices.

\end{appendices}
\bibliographystyle{IEEEtran}
\bibliography{reference.bib}

\vspace{12pt}

\end{document}